# Integrated and Adaptive Guidance and Control for Endoatmospheric Missiles via Reinforcement Meta-Learning

Brian Gaudet*
*University of Arizona, 1127 E. Roger Way, Tucson Arizona, 85721*

Roberto Furfaro†
*University of Arizona, 1127 E. Roger Way, Tucson Arizona, 85721*

**We apply a reinforcement meta-learning framework to optimize an integrated and adaptive guidance and flight control system for an air-to-air missile. The system is implemented as a policy that maps navigation system outputs directly to commanded rates of change for the missile's control surface deflections. The system induces intercept trajectories against a maneuvering target that satisfy control constraints on fin deflection angles, and path constraints on look angle and load. We test the optimized system in a six degrees-of-freedom simulator that includes a non-linear radome model and a strapdown seeker model, and demonstrate that the system adapts to both a large flight envelope and off-nominal flight conditions including perturbation of aerodynamic coefficient parameters and flexible body dynamics. Moreover, we find that the system is robust to the parasitic attitude loop induced by radome refraction and imperfect seeker stabilization. We compare our system's performance to a longitudinal model of proportional navigation coupled with a three loop autopilot, and find that our system outperforms this benchmark by a large margin. Additional experiments investigate the impact of removing the recurrent layer from the policy and value function networks, and performance with an infrared seeker.**

## I. Introduction

Design of guidance and control (G&C) systems for supersonic air-to-air missiles is complicated by factors that include the large flight envelope, a changing center of mass during rocket burn, flexible body dynamics, mismatches between the design and deployment environments, control saturation, and the inability to measure altitude, speed, and angle of attack. Moreover, the combination of look angle dependent radome refraction, imperfect seeker stabilization, and rate gyro bias results in a false indication of target motion, which gives rise to a parasitic attitude loop that reduces accuracy and can potentially destabilize the G&C system [1–3]. Current practice in air-to air-missile G&C treats guidance and control as separately optimized systems. For example, the guidance system might map the line of sight (LOS) rotation rate and closing speed to a commanded acceleration. This commanded acceleration is then mapped to commanded control surface deflections by the flight control system (FCS). This control system, often referred to as the missile autopilot, is implemented as three separate controllers for a skid to turn implementation: a roll stabilizer, a pitch controller, and a yaw controller [4]. The roll stabilizer drives the missile roll rate to zero, whereas the pitch and yaw controllers are each implemented as a three loop autopilot. In a given control channel (pitch or yaw), the three loop autopilot tracks the commanded acceleration from the guidance system by mapping the acceleration tracking error and rotational velocity to fin deflections. The response of these controllers is dependent on dynamic pressure, and missiles operating over a large flight envelope typically require some form of gain scheduling. As it is not practical to measure altitude and speed during missile flight, these gains are selected based on the predicted trajectory of the interceptor taking into account the initial conditions of the engagement scenario [4], and are fixed for the duration of the intercept. However, gain scheduling results in performance degradation when target maneuvers cause the actual trajectory to differ significantly from the predicted trajectory. An alternative is an adaptive autopilot that automatically adjusts autopilot parameters in a way that maximizes some performance metric [4].

One solution to improve the performance of a G&C system is to integrate the guidance and control subsystems, allowing the integrated system to exploit synergies between the guidance and control functions. There are currently two approaches commonly applied to integrated G&C. In the single loop approach, the G&C system maps the output of the state estimation and filtering system directly to actuator commands. In contrast, the two loop implementation uses separate guidance and flight control loops, but the guidance system takes into account the dynamics of the FCS, the response of the vehicle under varying aerodynamic regimes, path constraints, and actuator limits. There is currently no consensus in the research community as to which approach achieves the best performance. Advantages of the two loop approach include the ability to run the guidance loop at a lower rate than the FCS, and the ability to add a bias term to the guidance policy's commanded acceleration, with the bias term being a function of the estimated target acceleration. On the other hand, the single loop approach does not require spectral separation between multiple loops, allowing use of a lower effective guidance time constant, and reducing flight control response time. Importantly, the spectral separation between separate guidance and flight control systems might not be valid at the end of the engagement, where rapid changes in engagement geometry occur [5].

*Research Engineer, Department of Systems and Industrial Engineering, E-mail: briangaudet@arizona.edu

†Professor, Department of Systems and Industrial Engineering, Department of Aerospace and Mechanical Engineering. E-mail: robertof@arizona.edu

Recent work studying integrated G&C includes [5] where the authors develop a longitudinal integrated G&C system using continuous time predictive control, with the engagement state and disturbances estimated using a generalized extended state observer. A partially integrated (two loop) guidance and control system is developed in [6] using sliding mode control, and in [7] the authors use a back-stepping technique to develop an integrated system satisfying impact angle constraints. Most work to date in integrated G&C have used a longitudinal model, ignoring the effects of cross-coupling between the pitch and yaw channels [8]. One of the few papers to handle the full 6-DOF case is [9], where the authors use a two loop approach for a partially integrated G&C system in a surface to air missile application. Although there are two loops, the system is partially integrated in that the guidance loop is aware of the 6-DOF dynamics. Unfortunately, the guidance loop requires not only the full engagement state to be estimated, but also aerodynamic coefficients, and performance was not tested with uncertainty on these variables. Moreover, a radome model was not included in the simulation, so it is not clear if the parasitic attitude loop would destabilize the system, and the target does not attempt evasive maneuvers. An integrated and adaptive G&C system is developed in [10] using a longitudinal model with simplified non-linear dynamics and $\mathcal{L}^1$ adaptive control. It is worth noting that a single loop integrated and adaptive G&C system for an exo-atmospheric interceptor was developed in [11].

To date, research into integrated G&C for endoatmospheric missile applications has not taken advantage of recent advancements in deep learning methods and algorithms. Meta-RL has been demonstrated to be effective in optimizing integrated and adaptive G&C systems that generate direct closed-loop mapping from navigation system outputs to actuation commands. Applications where meta-RL has been effectively applied to GN&C include asteroid close proximity operations [12, 13], planetary landing [14–16], exoatmospheric intercept [11], and hypersonic vehicle guidance [17, 18]. In the meta-RL framework, an agent instantiating a policy learns how to complete a task through episodic simulated experience over a continuum of environments. The policy is implemented as a deep neural network that maps observations to actions $\mathbf{u} = \pi_\theta(\mathbf{o})$, and in our work is optimized using a customized version of proximal policy optimization (PPO)[19]. Adaptation is achieved by including a recurrent network layer with hidden state $\mathbf{h}$ in both the policy and value function networks. Maximizing the PPO objective function requires learning hidden layer parameters $\theta_\text{h}$ that result in $\mathbf{h}$ evolving in response to the history of $\mathbf{o}$ and $\mathbf{u}$ in a manner that facilitates fast adaptation to environments encountered during optimization, as well as novel environments outside of the training distribution. The optimized policy then adapts real-time to conditions experienced during deployment. Importantly, the network parameters remain fixed during deployment with adaptation occurring through the evolution of $\mathbf{h}$.

In this work we use the meta-RL framework to optimize an integrated and adaptive G&C system for an air-to-air missile application using a 6-DOF simulation environment. A system level view of the GN&C system is given in Fig. 1, where the policy replaces the traditional combination of a separate guidance and flight control system. Over the theater of operations defined by an engagement ensemble, the G&C system implements a closed loop mapping from navigation system outputs to commanded deflection rates for the missile's control surfaces. This mapping induces intercept trajectories that satisfy path constraints on load and look angle. Importantly, the system is optimized over an ensemble of aerodynamic models with variation in aerodynamic coefficients, resulting in a system that can adapt real time to differences in the optimization and deployment environments. Our simulator models look angle dependent radome refraction, imperfect LOS stabilization, and scale factor errors, all of which contribute to the parasitic attitude loop. In contrast to prior work, our G&C system uses only observations that are readily available from seeker and rate gyro outputs with minimal processing, and we model LOS stabilization. Moreover, we consider challenging target maneuvers, and we optimize and test the system using the full non-linear 6-DOF dynamics and an intuitive geometry based aerodynamics model derived from slender body and Newtonian impact theory. However, we decided to simplify the problem by not modeling the rocket boost phase.

The paper is organized as follows. In Section II we present the engagement scenarios, radome and rate gyro models,

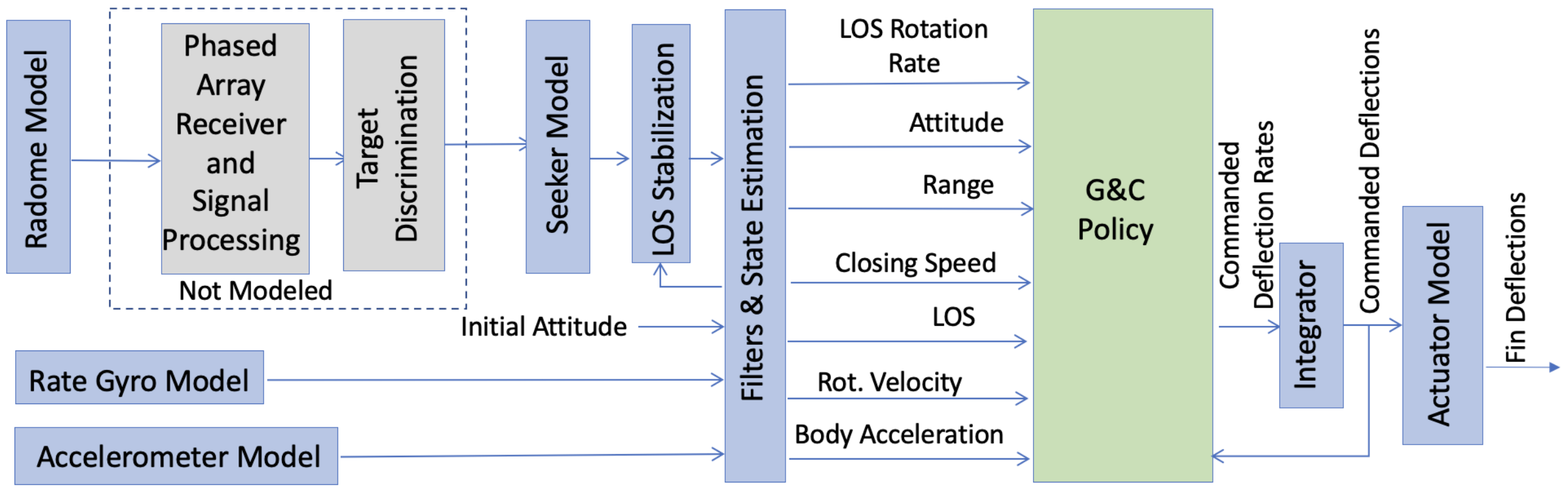

**Fig. 1 System Diagram of GN&C system**

actuator model, the aerodynamic model, and the equations of motion. Next in Section III.A we give a brief summary of the meta-RL framework. This is followed by Section III.B, where we formulate the meta-RL optimization problem for the air-to-air homing phase scenario described in Section II. In Section IV we optimize and test the integrated G&C policy, and then compare the results to a longitudinal benchmark using PN and a three loop autopilot, which is the standard for currently deployed missile systems. Additional experiments demonstrate the impact of removing the recurrent layer from the policy and value function networks, test the system with a flexible body dynamics model, and optimize and test a system with an infrared seeker, where no range and range rate measurements are available.

# II. Problem Formulation

## A. Engagement Geometry and Initial Conditions

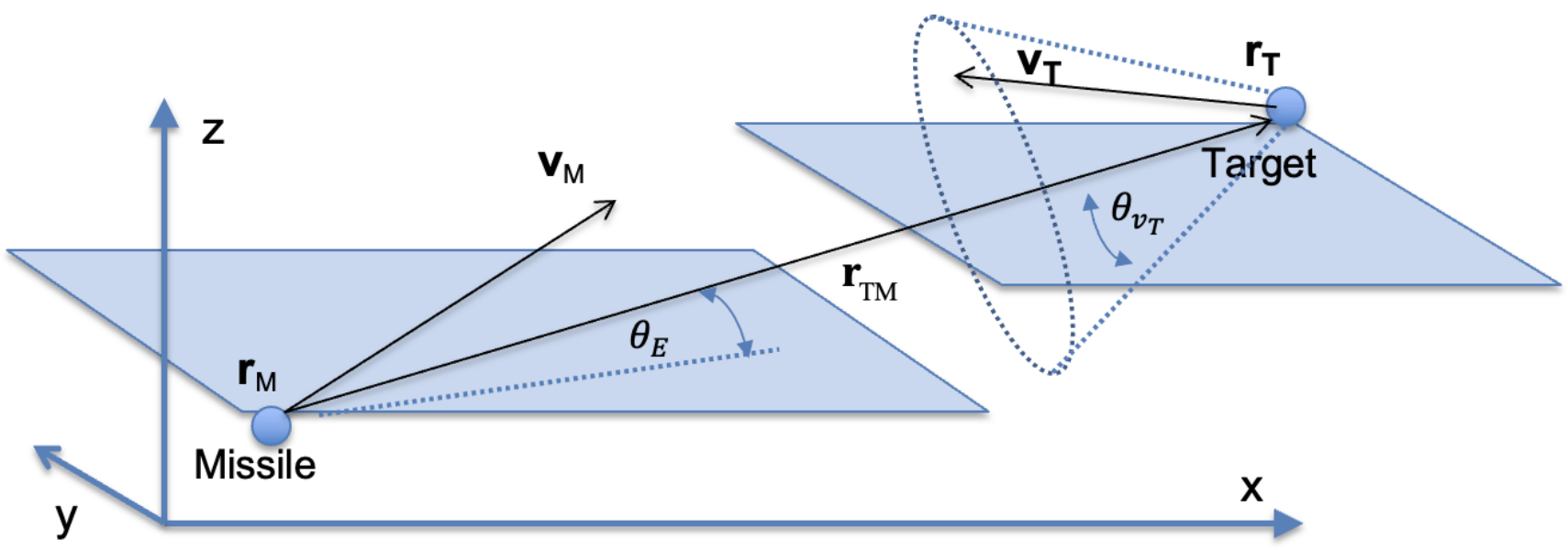

**Fig. 2 Engagement**

In this work we model a skewed head on engagement scenario. Due to the high target acceleration capability, which exceeds the level that a human pilot could withstand, a realistic scenario would be a fighter launching a missile to intercept a supersonic cruise missile that can actively maneuver to avoid threats. Referring to Fig. 2*, the missile position vector, missile velocity vector, target position vector, and target velocity vector are shown as $\mathbf{r}_{\mathrm{M}}$, $\mathbf{v}_{\mathrm{M}}$, $\mathbf{r}_{\mathrm{T}}$, $\mathbf{v}_{\mathrm{T}}$. We can also define the relative position and velocity vectors $\mathbf{r}_{\mathrm{TM}} = \mathbf{r}_{\mathrm{T}} - \mathbf{r}_{\mathrm{M}}$ and $\mathbf{v}_{\mathrm{TM}} = \mathbf{v}_{\mathrm{T}} - \mathbf{v}_{\mathrm{M}}$. The elevation angle $\theta_E$ is the angle between $\mathbf{r}_{\mathrm{TM}}$ and its projection onto the x-y plane.

We randomly generate the target's initial velocity vector such that $\mathbf{v}_{\mathrm{T}}$ lies within a cone with axis $\mathbf{r}_{\mathrm{TM}}$ and half apex angle $\theta_{\mathbf{v}_{\mathrm{T}}}$. A collision triangle is then defined in a plane that is not in general aligned with the coordinate frame shown in Fig. 2, and is illustrated in Fig. 3. Here we define the required lead angle $L$ for the missile's velocity vector $\mathbf{v}_{\mathrm{M}}$ as the angle that will put the missile on a collision triangle with the target in terms of the target velocity $\mathbf{v}_{\mathrm{T}}$, line-of-sight angle

*In this figure, the illustrated vectors are not in general within the x-z plane

$\gamma$, and the magnitude of the missile velocity as shown in Eqs. (1a) through (1c).

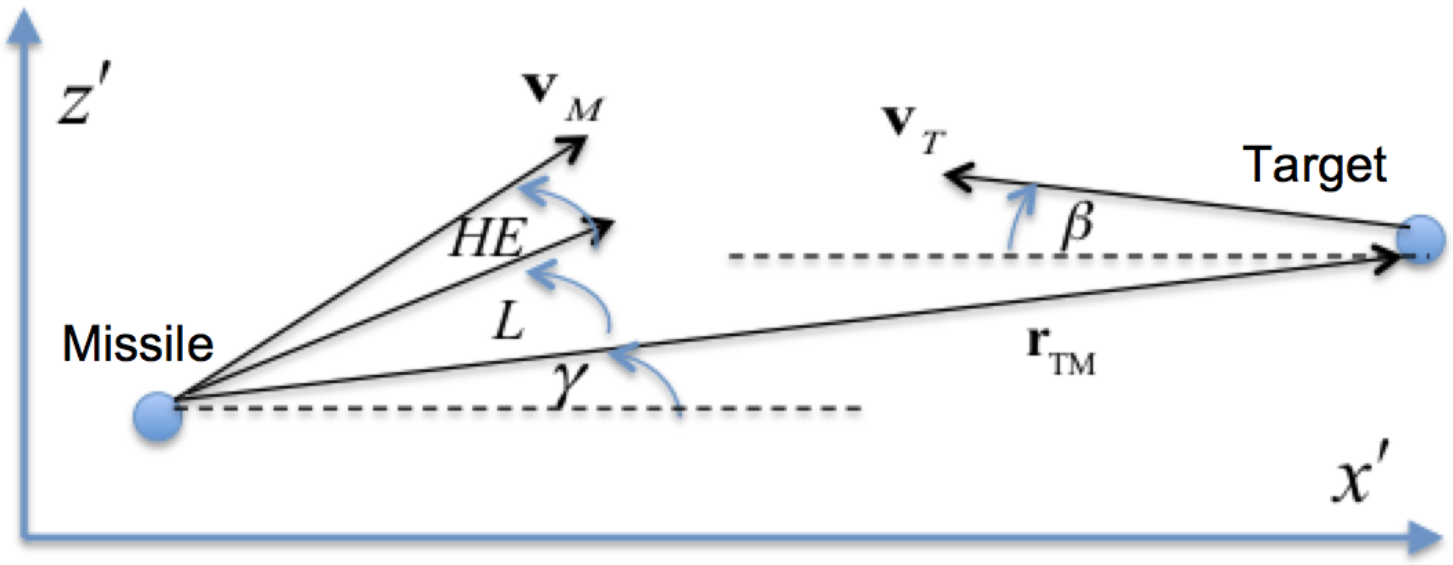

**Fig. 3 Planar Heading Error**

$$L = \arcsin\left(\frac{\|\mathbf{v}_\mathrm{T}\| \sin(\beta+\gamma)}{\|\mathbf{v}_\mathrm{M}\|}\right) \tag{1a}$$

$$v_{m_y} = \|\mathbf{v}_\mathrm{M}\| \cos(L+\gamma) \tag{1b}$$

$$v_{m_z} = \|\mathbf{v}_\mathrm{M}\| \sin(L+\gamma) \tag{1c}$$

This formulation is easily extended to a three dimensional engagement using the following approach:

1) define a plane normal as $\hat{\mathbf{v}}_t \times \hat{\boldsymbol{\lambda}}$
2) rotate $\mathbf{v}_\mathrm{T}$ and $\hat{\boldsymbol{\lambda}}$ onto the plane
3) calculate the required planar missile velocity (Eq. (1a))
4) rotate this velocity back into the original reference frame

Thus in $\mathbb{R}^3$ we define a heading error (HE) as the angle between the missile's initial velocity vector and the velocity vector associated with the lead angle required to put the missile on a collision heading with the target. Note that due to the missile aerodynamic forces and target acceleration, this is far from a perfect collision triangle, and the true heading error is greater than HE. Indeed, if we simulate with a zero heading error at a 10 km initial range, we observe a miss of 300m, indicating a true heading error of 1.7 degrees.

The simulator randomly chooses between a target bang-bang, weave, and jinking maneuver with equal probability, with the acceleration applied orthogonal to the target's velocity vector. The maneuvers have varying acceleration levels and random start time, duration, and switching time. At the start of each episode, with probability 0.5 the maneuvers in that episode use the target's maximum acceleration capability, and with probability 0.5 the acceleration is sampled uniformly between 0 and the maximum. We assume the target uses aerodynamic control surfaces (no thrust vector control). Consequently, the maximum target acceleration is reduced taking into account dynamic pressure. Specifically, we assume that the target can achieve the acceleration shown in Table 1 only at $q_o^{\mathrm{MAX}}$, the dynamic pressure corresponding to its maximum speed at sea level, but we reduce this maximum acceleration by the ratio of $\frac{q_o}{q_o^{\mathrm{MAX}}}$, where $q_o = \frac{1}{2}\rho\|\mathbf{v}_\mathrm{T}\|^2$. Sample target maneuvers are shown Fig. 4, note that in some cases the maneuver period is considerably shorter or longer, with the longest periods being twice the time of flight. We assume the target is powered and can maintain a constant speed during the maneuver. Note that although we optimize the G&C system with a missile acceleration advantage of 5:1, in Section IV we test the system with a missile acceleration advantage of up to 2:1.

We can now list the range of engagement scenario parameters in Table 1. During optimization and testing, these parameters are drawn uniformly between their minimum and maximum values, except as noted. The generation of heading error is handled as follows. We first calculate the optimal missile velocity vector that puts the missile on a collision triangle with the target as described previously. We then uniformly select a heading error $HE$ between the bounds given in Table 1, and randomly perturb the direction of the missile's velocity vector direction such that $\arccos(\mathbf{v}_\mathrm{M} \cdot \mathbf{v}_{\mathrm{M}_p}) < HE$, where $\mathbf{v}_{\mathrm{M}_p}$ is the perturbed missile velocity vector. Since the missile is launched by an aircraft, the initial angle of attack, side slip angle, and roll can vary as shown in Table 1. At the start of each episode, the aerodynamic force and moment coefficients described in Section II.F are independently perturbed by $\boldsymbol{\epsilon}_{\mathrm{force}}$ as shown in Table 4.

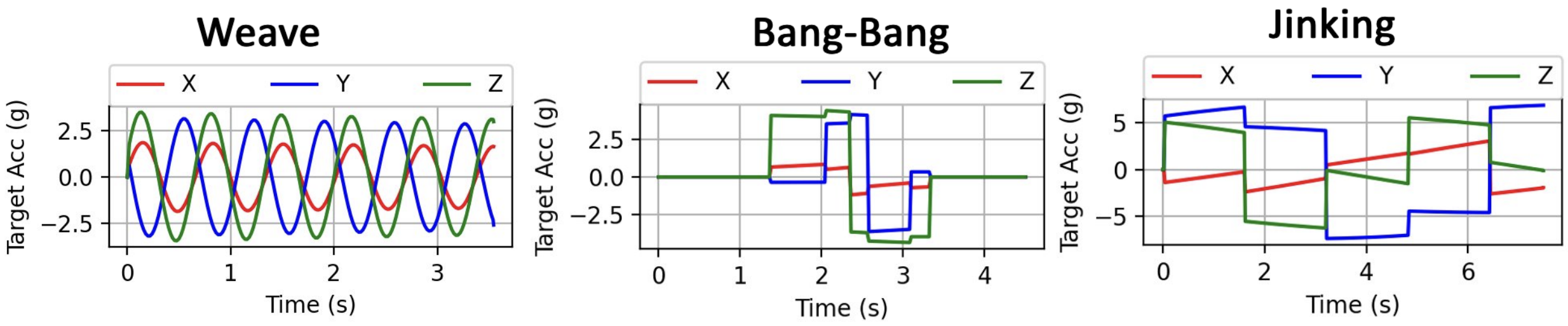

**Fig. 4 Sample Target Maneuvers**

**Table 1 Simulator Initial Conditions for Optimization**

| Parameters Drawn Uniformly | Min | Max |
|---|---|---|
| Range $\|\|\mathbf{r}_{\text{TM}}\|\|$ (m) | 5000 | 10000 |
| Elevation Angle $\theta_E$ (degrees) | -30 | 30 |
| Missile Velocity Magnitude $\|\|\mathbf{v}_{\text{M}}\|\|$ (m/s) | 800 | 1000 |
| Target Velocity Magnitude $\|\|\mathbf{v}_{\text{T}}\|\|$ (m/s) | 250 | 600 |
| Target Velocity Cone Half Apex Angle $\theta_{\mathbf{v}_{\text{T}}}$ (degrees) | 30 | 30 |
| Heading Error (degrees) | 0 | 5 |
| Initial Angle of Attack $\alpha$ (degrees) | -10 | 10 |
| Initial Side Slip Angle $\beta$ (degrees) | -5 | 5 |
| Initial Roll Angle (degrees) | -30 | 30 |
| Target Maximum Acceleration @ $q_o^{\text{MAX}}$ (m/s$^2$) | 0 | 8×9.81 |
| Target Bang-Bang duration (s) | 1 | 8 |
| Target Bang-Bang initiation time (s) | 0 | 6 |
| Target Weave Period (s) | 1 | 8 |
| Target Weave Offset (s) | 1 | 5 |

The missile G&C system is optimized to satisfy path constraints, these are tabulated in Table 2. The path constraints on attitude are not actually required, but reduced optimization run time by keeping the agent out of regions of unproductive state space. The minimum speed constraint is imposed to insure the missile's speed does not fall to subsonic speeds, in which case the calculation for $\eta = \sqrt{M^2 - 1}$ in Section II.F would not be correct. The look angle constraint implements a field of view constraint, this constrains the maximum angle between the body frame x-axis and the body frame LOS vector to be less than 45°.

**Table 2 Path Constraints**

| Constraint | Min | Max |
|---|---|---|
| Minimum Speed (m/s) | 400 | 400 |
| Pitch (degrees) | -85 | 85 |
| Yaw (degrees) | -85 | 85 |
| Roll (degrees) | -100 | 100 |
| X component of Rotational Velocity Vector $\omega_{\text{x}}$ (degrees / s) | -15 | 15 |
| Look angle $\theta_{\text{L}}$ (degrees) | - | 45 |
| Load $\|\|[F_y^{\text{B}}, F_z^{\text{B}}]\|\|$ (g) | - | 45 |

### B. Radome, LOS stabilization, Rate Gyro and Accelerometer Models

Let $\boldsymbol{\lambda}^{\text{N}} = \dfrac{\mathbf{r}_{\text{TM}}}{\|\mathbf{r}_{\text{TM}}\|}$ be the inertial frame line of sight (LOS) unit vector. The body frame LOS unit vector is calculated as $\boldsymbol{\lambda}^{\text{B}} = \mathbf{C}_{\text{BN}}\boldsymbol{\lambda}^{\text{N}}$, where $\mathbf{C}_{\text{BN}}$ is the direction cosine matrix (DCM) mapping from the inertial to body frame. Radome refraction will cause the apparent LOS $\tilde{\boldsymbol{\lambda}}^{\mathrm{B}}$ to differ from the ground truth LOS $\boldsymbol{\lambda}^{\mathrm{B}}$, and we define the radome refraction angle $\theta_R$ as the angle between $\boldsymbol{\lambda}^{\mathrm{B}}$ and $\tilde{\boldsymbol{\lambda}}^{\mathrm{B}}$, i.e., $\theta_R = \arccos(\boldsymbol{\lambda}^{\mathrm{B}} \cdot \tilde{\boldsymbol{\lambda}}^{\mathrm{B}})$.

We assume a symmetrical radome, where $\theta_R$ is a function of look angle $\theta_L = \arccos(\boldsymbol{\lambda}^{\mathrm{B}} \cdot \mathbf{e}_w^{\mathrm{B}})$, where $\mathbf{e}_w^{\mathrm{B}} = [1, 0, 0]$ is the missile centerline (body frame x-axis). First, we calculate the azimuthal ($\theta_u$) and elevation ($\theta_v$) refraction errors as shown in Eqs. (2a) through (2b), where $A_u$, $A_v$, $k_u$, and $k_v$ are sampled uniformly within the bounds given in Table 3 at the start of each simulation episode.

$$\theta_u = A_u \left(0.75\frac{\theta_L}{\pi/2} + 0.25\cos\left(\frac{2\pi}{k_u}\theta_L\right)\right) \tag{2a}$$

$$\theta_v = A_v \left(0.75\frac{\theta_L}{\pi/2} + 0.25\cos\left(\frac{2\pi}{k_v}\theta_L\right)\right) \tag{2b}$$

**Table 3 Radome Model Parameter Bounds**

| Variable | Lower limit | Upper Limit |
|---|---|---|
| $A_u$ | -1e-2 | 1e-2 |
| $A_v$ | -1e-2 | 1e-2 |
| $k_u$ | 1.00 | 3.00 |
| $k_v$ | 1.00 | 3.00 |

We then create the refracted body frame LOS unit vector $\tilde{\boldsymbol{\lambda}}^{\mathrm{B}} = \mathbf{C}(\mathbf{q}_{\mathrm{R}})\boldsymbol{\lambda}^{\mathrm{B}}$, where $\mathbf{C}(\mathbf{q}_{\mathrm{R}})$ is the DCM corresponding to the 321 Euler rotation $\mathbf{q}_{\mathrm{R}} = [\theta_u, \theta_v, 0]$. The LOS unit vector signal output by the seeker model is $\boldsymbol{\lambda}_{\mathrm{obs}}^{\mathrm{B}} = \mathbf{C}(\mathbf{q}_{\mathrm{N}})\tilde{\boldsymbol{\lambda}}^{\mathrm{B}}$, where $\mathbf{q}_{\mathrm{N}} = \mathcal{N}(\mu, \sigma_{\mathrm{LOS}}, 3)$ is a stochastic Euler 321 rotation (to model Gaussian white noise on the LOS measurement), and $\mathcal{N}(\mu, \sigma, n)$ denotes an $n$ dimensional normally distributed random variable with mean $\mu$ and standard deviation $\sigma$. We use $\sigma_{\mathrm{LOS}} = 1$ mrad.

The resulting refraction angle $\theta_R$ is shown in Fig. 5 for the case of $A_u = A_v = 10$ mrad and various values of $k$. Note that the radome slope $\frac{\partial\theta_R}{\partial\theta_L}$ is given by the slope of the curves in the figure.

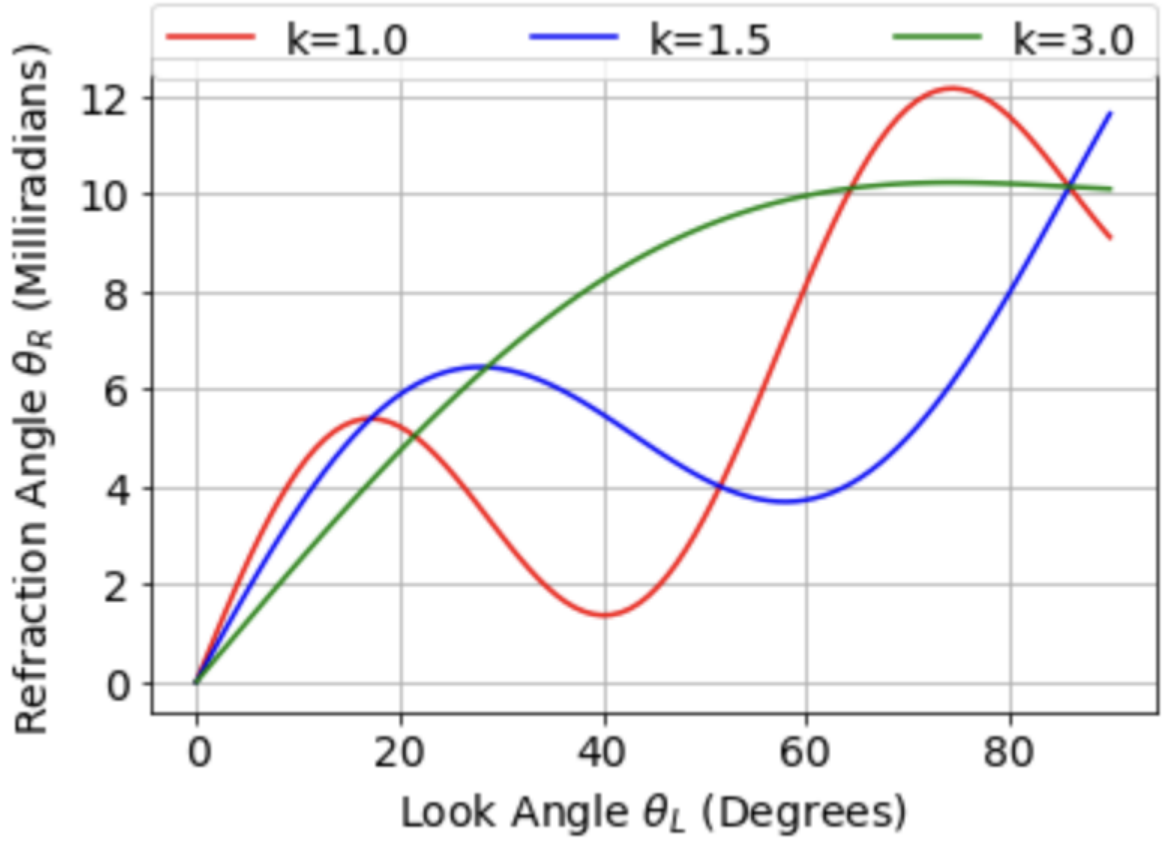

**Fig. 5 Radome Refraction Angle as Function of Look Angle**

In addition, the seeker model outputs a closing speed measurement $v_{c_{\mathrm{obs}}} = -\frac{\mathbf{r}_{\mathrm{TM}}^{\mathrm{B}} \cdot \mathbf{v}_{\mathrm{TM}}^{\mathrm{B}}}{\|\mathbf{r}_{\mathrm{TM}}^{\mathrm{B}}\|}$ and a range measurement $r_{\mathrm{obs}} = \|\mathbf{r}_{\mathrm{TM}}^{\mathrm{B}}\|$.

Our rate gyro model corrupts the ground truth rotational velocity $\boldsymbol{\omega}$ with both scale factor bias and Gaussian noise, as shown in Eq. (3a), where $\mathcal{U}(a, b, n)$ denotes an $n$ dimensional uniformly distributed random variable bounded by $(a, b)$ that is generated at the start of each episode, and each dimension of the random variable is independent. We use $\epsilon_\omega = 0.001$ and $\sigma_\omega = 0.001$.

$$\omega^{\mathrm{B}}_{\mathrm{obs}} = \omega(1 + \mathcal{U}(-\epsilon_\omega, \epsilon_\omega, 3)) + \mathcal{N}(0, \sigma_\omega, 3) \tag{3a}$$

We found that providing the meta-RL G&C system with an estimated body frame acceleration improved satisfaction of the load constraint. Similar to the rate gyro model, the observed body frame acceleration is corrupted with a scale factor error as shown in Eq. (4a), where we use $\epsilon_{\mathrm{acc}} = 0.001$.

$$\mathbf{a}^{\mathrm{B}}_{\mathrm{obs}} = \mathbf{a}^{\mathrm{B}}(1 + \mathcal{U}(-\epsilon_{\mathrm{acc}}, \epsilon_{\mathrm{acc}}), 3))) \tag{4a}$$

### C. LOS Stabilization

Referring to Fig. 1, the refracted body frame line of sight unit vector $\boldsymbol{\lambda}^{\mathrm{B}}_{\mathrm{obs}}$ is input to the LOS stabilization block. This block uses the estimated change of attitude since the start of the engagement $\mathbf{dq}$ to rotate $\boldsymbol{\lambda}^{\mathrm{B}}_{\mathrm{obs}}$ into the reference frame corresponding to the interceptor's attitude at the start of the engagement. Specifically, at each guidance step we map $\mathbf{dq}$ to a direction cosine matrix $\mathbf{C}(\mathbf{dq})$, and compute the stabilized LOS as $\boldsymbol{\lambda}^{\mathrm{S}}_{\mathrm{obs}} = [\mathbf{C}(\mathbf{dq})]^{\mathrm{T}}\boldsymbol{\lambda}^{\mathrm{B}}_{\mathrm{obs}}$.

The estimated change in attitude $\mathbf{dq}_{\mathrm{obs}}$ is parameterized as a quaternion, and is computed by integrating $\boldsymbol{\omega}$ (from the filtering block described in Section II.D) as shown in Equation 5, where $\mathbf{dq}_{\mathrm{obs}}$ is reset at the start of each episode to $\mathbf{dq}_{\mathrm{obs}_0} = \begin{bmatrix} 1 & 0 & 0 & 0 \end{bmatrix}$. In our simulation model, we approximate this integration using fourth order Runge-Kutta integration. In the unrealistic case where we use the ground truth value of $\boldsymbol{\omega}$, this completely decouples $\boldsymbol{\lambda}^{\mathrm{S}}$ from body rotation. However, since $\boldsymbol{\omega}$ is corrupted with a scale factor error and Gaussian noise, in general $\mathbf{dq}_{\mathrm{obs}} \neq \mathbf{dq}$, and the LOS stabilization will be imperfect. The combination of imperfect stabilization and radome refraction result in a false indication of target motion. This results in increased miss distance, and can potentially destabilize the G&C system. This is discussed in more detail in [1, 2, 11].

$$\begin{bmatrix} \dot{dq0} \\ \dot{dq1} \\ \dot{dq2} \\ \dot{dq3} \end{bmatrix} = \frac{1}{2} \begin{bmatrix} dq0 & -dq1 & -dq2 & -dq3 \\ dq1 & dq0 & -dq3 & dq2 \\ dq2 & dq3 & dq0 & -dq1 \\ dq3 & -dq2 & dq1 & dq0 \end{bmatrix} \begin{bmatrix} 0 \\ \omega_0 \\ \omega_1 \\ \omega_2 \end{bmatrix} \tag{5}$$

### D. Filtering and State Estimation

The LOS rotation rate is calculated as shown in Eq. (6).

$$\boldsymbol{\Omega} = \frac{r_{\mathrm{obs}}\boldsymbol{\lambda}^{\mathrm{S}}_{\mathrm{obs}} \times \mathbf{v}^{\mathrm{S}}_{\mathrm{TM}}}{r^2_{\mathrm{obs}}} \tag{6}$$

However, $\mathbf{v}^{\mathrm{S}}_{\mathrm{TM}}$ is not directly measurable from sensor outputs, although it can be estimated along with $\mathbf{r}^{\mathrm{S}}_{\mathrm{TM}}$ using a non-linear Kalman filter [20]. Indeed, a single Kalman filter can both filter noise from $\boldsymbol{\lambda}^{\mathrm{S}}_{\mathrm{obs}}$ and infer $\mathbf{v}^{\mathrm{S}}_{\mathrm{TM}}$. The filter's state propagation model would model target acceleration as a random walk (its derivative equal to white noise), and the measurement model would use $\boldsymbol{\lambda}^{S}_{\mathrm{obs}}$, $r_{\mathrm{obs}}$, and $v_{c_{\mathrm{obs}}}$. Since $r_{\mathrm{obs}}$, and $v_{c_{\mathrm{obs}}}$ are non-linear functions of the state variables, this requires either an extended or unscented Kalman filter implementation.

In this work we do not model the Kalman filter, but instead pass $\boldsymbol{\lambda}^{\mathrm{S}}_{\mathrm{obs}}$ through a single pole low pass filter. Further, the ground truth value of $\mathbf{v}_{\mathrm{TM}}$ is corrupted with Gaussian noise with a standard deviation of 1 mrad and passed through an additional low pass filter before it is used in the calculation of $\boldsymbol{\Omega}$. We also pass $\boldsymbol{\omega}_{\mathrm{obs}}$ through a low pass filter. We use a navigation frequency of 100 Hz, i.e., signals from the radome, seeker, rate gyro, and accelerometer models are sampled every 0.01 s. Low pass filters use a time constant of 0.01s.

In an alternate approach used in the experiments (Section IV), we use a surrogate for the line of sight rotation rate that does not require estimation of $\mathbf{v}_{\mathrm{TM}}$, and could be implemented on an infrared seeker that does not measure range or range rate. This surrogate LOS rate is calculated as shown in Eq. (7).

$$\mathbf{\Omega}_{\mathrm{surr}} = \frac{\boldsymbol{\lambda}^{\mathrm{S}}_{\mathrm{obs}_t} \times \boldsymbol{\lambda}^{\mathrm{S}}_{\mathrm{obs}_{t-\Delta t}}}{\Delta t} \tag{7}$$

The intuition for the surrogate rotation rate is that the cross product increases with the angle between the two samples of $\boldsymbol{\lambda}$. Although we have not rigorously analyzed the surrogate rotation rate, inspection of Eqs. (6) and (7) show that both the surrogate and correct rotation rate vectors are orthogonal to the LOS vector.

The missile attitude observation is computed as $\mathbf{q}_{\mathrm{obs}} = \mathbf{q}_{\mathrm{init}} \circ \mathbf{dq}$, where $\mathbf{q}_{\mathrm{init}}$ is the missile attitude at launch, which we assume can be set accurately by the launching aircraft.

### E. Actuator Model

We found that meta-RL optimization efficiency is improved by interpreting policy actions as commanded control rates, which leads to smoother changes in control surface deflections. The output of the guidance policy $\mathbf{u} = \pi(\mathbf{o}) \in \mathbb{R}^4$ described in Section III.B is split into four scalar components, a commanded deflection rate to apply to both the missile's upper and lower tail control surfaces, a deflection rate to apply to both the missile's left and right tail control surfaces, a differential deflection rate to apply to both the missile's upper and lower tail control surfaces, and a differential deflection rate to apply to both the missile's left and right tail control surfaces. The combined and differential commanded deflection rates are generated by scaling the associated element of $\mathbf{u}$ with the maximum allowed commanded rate, after which the elements are clipped to fall within the allowed commanded deflection rates. For combined and differential deflections, the limits are 80°/s and 1°/s.

These are then integrated to obtain the commanded combined and differential deflections, and clipped to fall between the combined and differential actuator limits of 20° and 0.1°. These commanded deflections are then transformed to full commanded fin deflections. i.e., $\theta_{\mathrm{LT}} = \theta_{\mathrm{H}} - \Delta\theta_{\mathrm{H}}$, $\theta_{\mathrm{RT}} = \theta_{\mathrm{H}} + \Delta\theta_{\mathrm{H}}$, $\theta_{\mathrm{BT}} = \theta_{\mathrm{V}} - \Delta\theta_{\mathrm{V}}$, and $\theta_{\mathrm{UT}} = \theta_{\mathrm{V}} + \Delta\theta_{\mathrm{V}}$, where $\theta_{\mathrm{H}}$ and $\theta_{\mathrm{V}}$ are the combined horizontal and vertical deflections, and $\Delta\theta_{\mathrm{H}}$ and $\Delta\theta_{\mathrm{V}}$ are the differential horizontal and vertical deflections. The full commanded fin deflections $\theta_{\mathrm{defl}}$ are then passed to the actuator dynamics. For meta-RL optimization the actuator dynamics is implemented as a first order lag with $\tau = 0.01$s, but we test the optimized system using the second order actuator dynamics model suggested in [21], with the transfer function shown in Eq. (8) applied to each deflection using $\zeta_{\mathrm{ACT}} = 0.7$ and $\omega_{\mathrm{ACT}} = 600$ rad/s. The output of the actuator dynamics are the control surface deflections $[\theta_{LT}, \theta_{RT}, \theta_{BT}, \theta_{UT}]$ that are input to the aerodynamics model described in Section II.F. Note that the reason we do not optimize with the second order model is due to the high frequency pole, which necessitates reducing the integration step size to a value that would significantly slow optimization.

$$\frac{1}{1 + \frac{2\zeta_{\mathrm{ACT}}}{\omega_{\mathrm{ACT}}} s + \frac{s^2}{\omega^2_{\mathrm{ACT}}}} \tag{8}$$

### F. Aerodynamic Model

Our aerodynamic model is derived from slender body and Newtonian impact theory, and we use the same missile geometry as described in [22], which allows us to compare our results to the longitudinal benchmark given in [23]. The missile geometry is shown in Fig. 6, which is drawn only approximately to scale. Here $d = 0.31$, $h_{\mathrm{W}} = 0.63$, $c_{\mathrm{RW}} = 1.88$, $c_{\mathrm{RT}} = 0.63$, $h_{\mathrm{T}} = 0.63$, $h_{\mathrm{W}} = 0.63$, $x_{\mathrm{HL}} = 6.09$, $x_{\mathrm{L}} = 6.25$, $x_{\mathrm{CG}} = 3.13$, and $x_{\mathrm{N}} = 0.94$, with all values in meters. Our approach is similar to Zarchan's longitudinal model [22], but extended to 6-DOF, including the modeling of roll coupling. In this work we model the portion of the homing phase that occurs after rocket burnout, so the center of gravity remains constant.

The body frame x-component of the center of pressure locations are combined into a vector as shown in Eq. (9a), where the vector components correspond to the nose, wing, body, and tail, respectively. To model the rolling moments, we also require the $y$ component of the left and right tail centers of pressure $\mathrm{cplt}_y$ and $\mathrm{cprt}_y$, as well as the $z$ component of the lower and upper tail centers of pressure $\mathrm{cpbt}_z$ and $\mathrm{cput}_z$; these are shown in Eqs. (9b) and (9c).

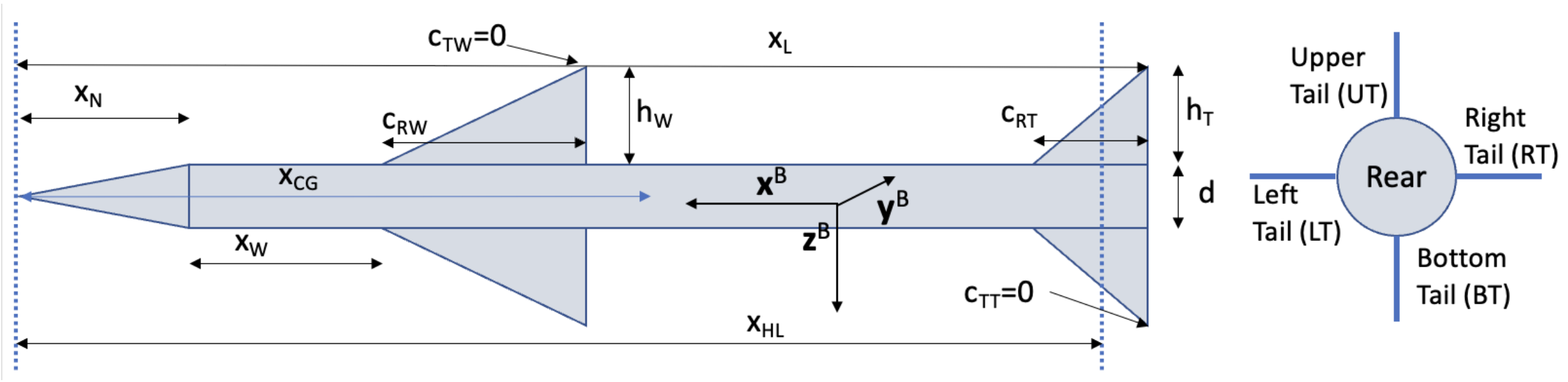

**Fig. 6 Missile Geometry**

$$\mathbf{cp}_x = \begin{bmatrix} 0.67x_{\mathrm{N}} & x_{\mathrm{N}} + x_{\mathrm{W}} + 0.7c_{\mathrm{RW}} - 0.2c_{\mathrm{TW}} & 0.67a_{\mathrm{N}}x_{\mathrm{N}} + a_{\mathrm{B}}\dfrac{x_{\mathrm{N}} + 0.5(x_{\mathrm{L}} - x_{\mathrm{n}})}{a_{\mathrm{N}} + a_{\mathrm{B}}} & x_{\mathrm{HL}} \end{bmatrix} \tag{9a}$$

$$\mathrm{cplt}_y = \mathrm{cpbt}_z = -(h_{\mathrm{W}}/2 + d/2) \tag{9b}$$

$$\mathrm{cprt}_y = \mathrm{cput}_z = h_{\mathrm{W}}/2 + d/2 \tag{9c}$$

The normal and side force coefficient vectors are given in Eqs. (10a) and (10b), where the vector components correspond to the contribution from the nose, wing, body, and tail, respectively.

$$\mathbf{C}_N = \begin{bmatrix} \sin 2\alpha & \dfrac{8S_{\mathrm{WING}}\sin\alpha}{\eta S_{\mathrm{REF}}} & \dfrac{1.5S_{\mathrm{PLAN}}\mathrm{sign}(\alpha)(\sin\alpha)^2}{\eta S_{\mathrm{REF}}} & \dfrac{8S_{\mathrm{TAIL}}\sin(\alpha - \theta_{LT})}{2\eta S_{\mathrm{REF}}} + \dfrac{8S_{\mathrm{TAIL}}\sin(\alpha - \theta_{RT})}{2\eta S_{\mathrm{REF}}} \end{bmatrix} \tag{10a}$$

$$\mathbf{C}_Y = \begin{bmatrix} \sin 2\beta & \dfrac{8S_{\mathrm{WING}}\sin\beta}{\eta S_{\mathrm{REF}}} & \dfrac{1.5S_{\mathrm{PLAN}}\mathrm{sign}(\beta)(\sin\beta)^2}{\eta S_{\mathrm{REF}}} & \dfrac{8S_{\mathrm{TAIL}}\sin(\beta - \theta_{BT})}{2\eta S_{\mathrm{REF}}} + \dfrac{8S_{\mathrm{TAIL}}\sin(\beta - \theta_{UT})}{2\eta S_{\mathrm{REF}}} \end{bmatrix} \tag{10b}$$

The axial force coefficient $C_A$ is given as shown in Eq. (11a); we use $k = 4$ and $C_{A_0} = 0.35$, where the second term gives the axial force component due to tail deflections, which induce a normal force, side force, or both.

$$C_A = C_{A_0} + k\left\|\begin{bmatrix} \sum \mathbf{C}_N & \sum \mathbf{C}_Y \end{bmatrix}\right\| \tag{11a}$$

The pitching, yawing moments, and rolling are computed as shown in Eqs. (12a) through (12e).

$$\mathbf{C}_m = \frac{\mathbf{C}_N(x_{\mathrm{CG}} - \mathbf{cp}_x)}{d} \tag{12a}$$

$$\mathbf{C}_n = \frac{\mathbf{C}_Y(x_{\mathrm{CG}} - \mathbf{cp}_x)}{d} \tag{12b}$$

$$C_{l_h} = -\left(\mathrm{cg}_y - \mathrm{cplt}_y\right)\frac{8S_{\mathrm{TAIL}}\sin(\alpha - \theta_{LT})}{2\eta S_{\mathrm{REF}}} - \left(\mathrm{cg}_y - \mathrm{cprt}_y\right)\frac{8S_{\mathrm{TAIL}}\sin(\alpha - \theta_{RT})}{2\eta S_{\mathrm{REF}}} \tag{12c}$$

$$C_{l_v} = \left(\mathrm{cg}_z - \mathrm{cpbt}_y\right)\frac{8S_{\mathrm{TAIL}}\sin(\beta - \theta_{BT})}{2\eta S_{\mathrm{REF}}} + \left(\mathrm{cg}_z - \mathrm{cput}_z\right)\frac{8S_{\mathrm{TAIL}}\sin(\beta - \theta_{UT})}{2\eta S_{\mathrm{REF}}} \tag{12d}$$

$$C_l = C_{l_h} + C_{l_v} \tag{12e}$$

The body frame force $\mathbf{F}^{\mathrm{B}}$ and torque $\mathbf{L}^{\mathrm{B}}$ are then computed as given in Eqs. (13a) through (13b). Here $C_{l_{\mathrm{damp}}} = -5$ is the roll damping coefficient, $C_{l_\beta} = 0.1$ is the coupling coefficient between sideslip angle and roll, $C_{l_\alpha} = 0.1$ is the coupling coefficient between angle-of-attack and roll, and $V$ is the missile speed.

$$\mathbf{F}^{\mathrm{B}} = qS_{\mathrm{REF}}\begin{bmatrix}-C_A & -\sum \mathbf{C}_Y & -\sum \mathbf{C}_N\end{bmatrix} \tag{13a}$$

$$\mathbf{L}^{\mathrm{B}} = qS_{\mathrm{REF}}d\begin{bmatrix}\left(C_{l_{\mathrm{damp}}}\dfrac{\mathrm{rad2deg}(\omega[0])d}{2V} + C_{l_\alpha} + C_{l_\beta} + C_l\right) & \sum \mathbf{C}_m & -\sum \mathbf{C}_n\end{bmatrix} \tag{13b}$$

A plot illustrating lift to drag (normal and axial transformed to the wind frame) are given in Fig. 7. This agrees reasonably well with the lift to drag of a cylinder with tapered nose with a length to diameter ratio of 10 as shown in [24] Figs 8 and 9, but our lift to drag is lower due to our length to diameter ratio of 20. Figure 8 plots the open loop response to a fin deflection [LT, RT, BT, UT] = (4.9, 5.1, 5.1, 4.9) at sea level, and Fig. 9 plots the open loop response to a fin deflection [LT, RT, BT, UT] = (19.9, 20.1, 20.1, 19.9) at an altitude of 15km. The rotational velocity initially is driven negative due to the angle of attack and sideslip angle to roll coupling, but the differential deflection then compensates, driving the roll rate positive. The plotted acceleration is in the missile body frame. The high altitude open loop response appears to be only marginally stable.

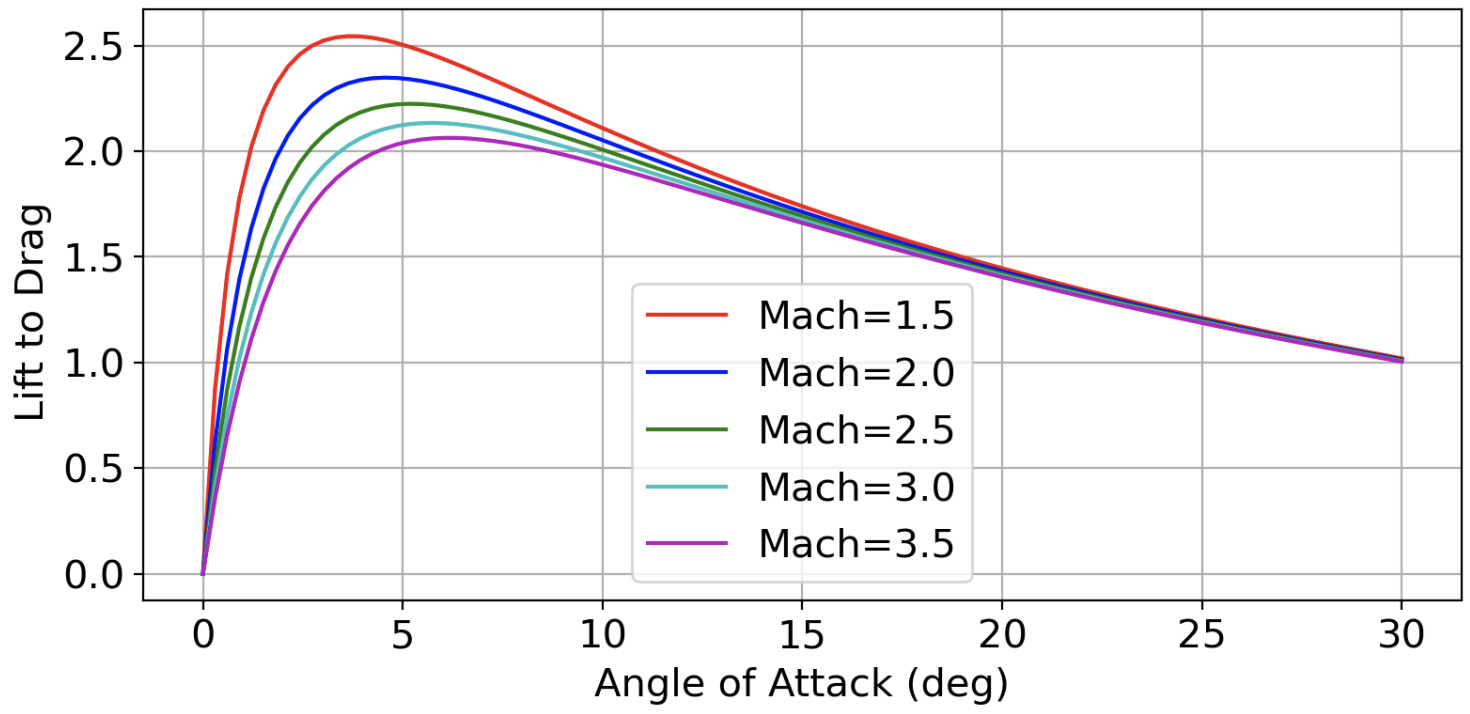

**Fig. 7 Lift versus Drag**

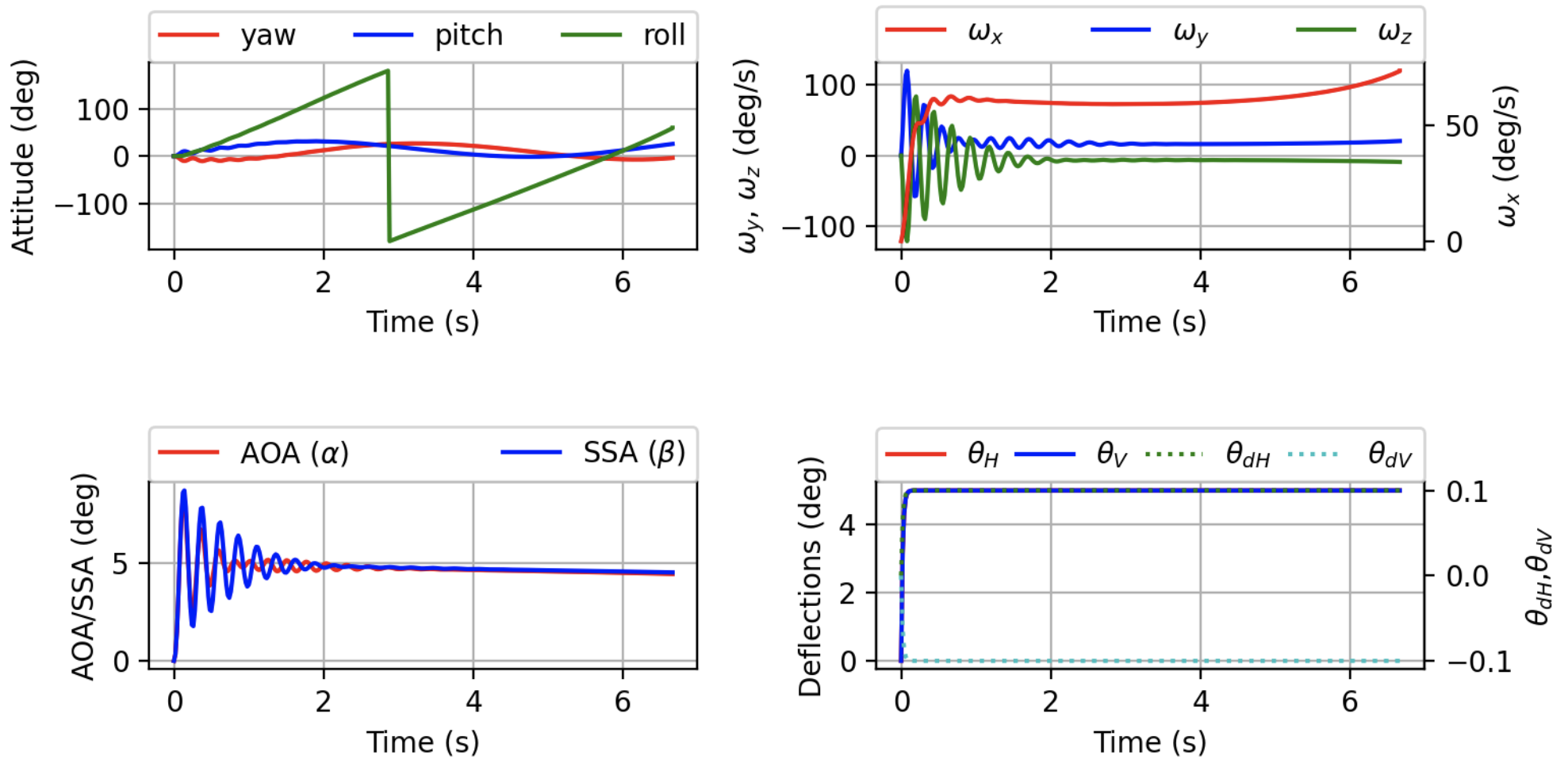

**Fig. 8 Open Loop Airframe Response to Fin Deflection, Altitude = 0 km**

We approximate the missile as a cylinder for purposes of computing the missile's inertia tensor, which is given in Figure 14, where $r = d/2$, and $m = 455$ kg is the missile mass.

$$\mathbf{J} = m\begin{bmatrix} r^2/2 & 0 & 0 \\ 0 & (3r^2 + x_{\mathrm{L}}^2)/12 & 0 \\ 0 & 0 & (3r^2 + x_{\mathrm{L}}^2)/12 \end{bmatrix} \tag{14}$$

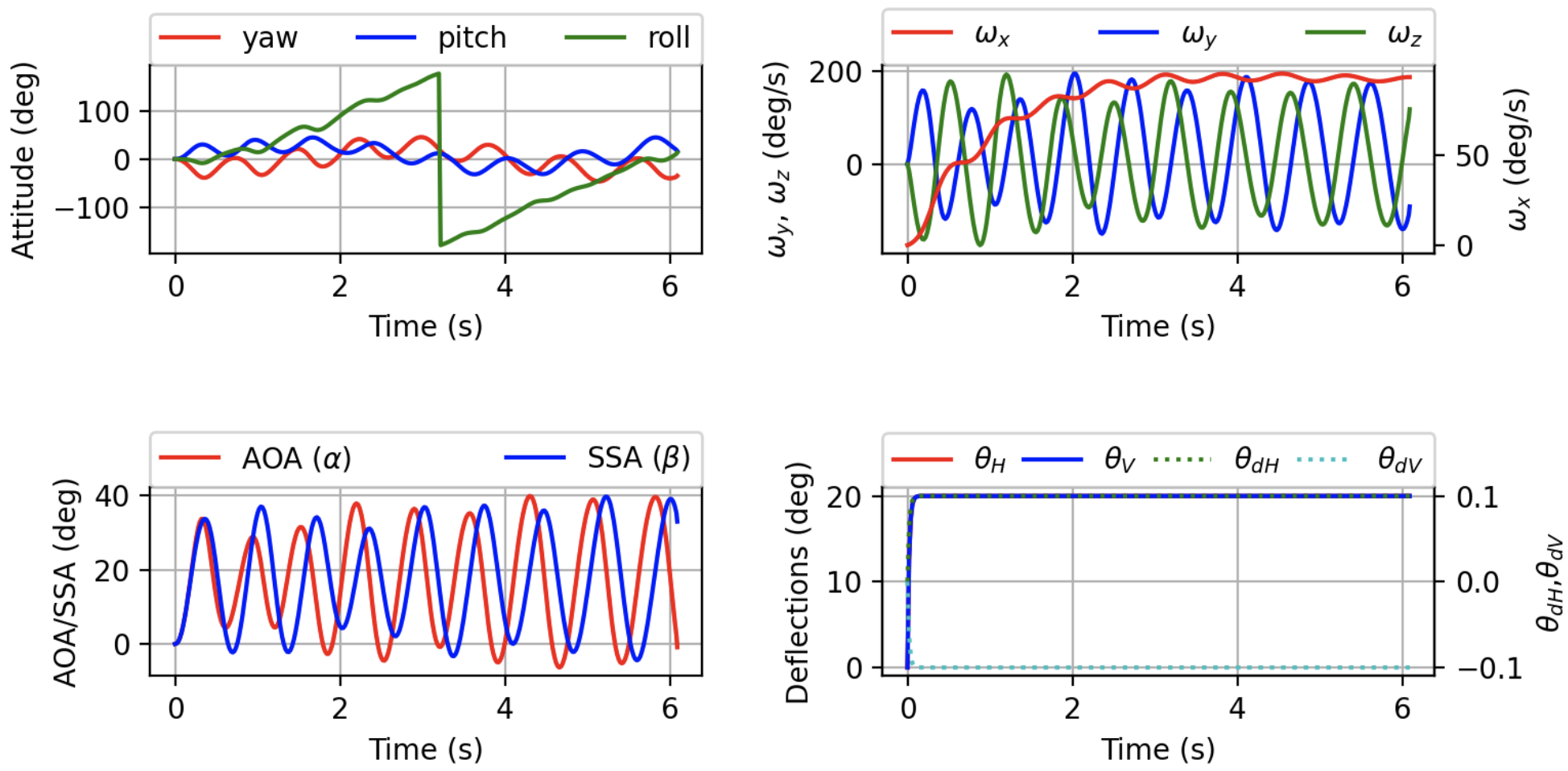

**Fig. 9 Open Loop Airframe Response to Fin Deflection, Altitude = 15 km**

In order to model differences between the optimization aerodynamics and deployment aerodynamics, at the start of each episode we multiply the force and moment coefficients by the corresponding element of $(1 + \boldsymbol{\epsilon}_{\text{coeff}})$, where with probability $p = 0.5$ a given episode has $\boldsymbol{\epsilon}_{\text{coeff}} \in \mathbb{R}^6$ uniformly drawn from the bounds show in Table 4, otherwise a coefficient is randomly set to $1 \pm \boldsymbol{\epsilon}_{\text{coeff}}$ for the duration of an episode, noting that in both cases each component of $\epsilon$ is independent.

**Table 4 Aerodynamic Coefficient Perturbation**

| Parameters Drawn Uniformly with probability 0.5 and randomly set to +/- Max otherwise | min | max |
|---|---|---|
| Aerodynamic Coefficient Variation $\boldsymbol{\epsilon}_{\text{coeff}}$ | -0.1 | 0.1 |

### G. Equations of Motion

Let $\mathbf{v}^{\text{B}} \equiv [u, v, w]$ denote the body frame missile velocity. The angle of attack $\alpha$ and sideslip angle $\beta$ are then computed as shown in Eqs. (15a) through (15b).

$$\alpha = \arctan \frac{w}{u} \tag{15a}$$

$$\beta = \arcsin \frac{v}{\|\mathbf{v}^{\text{B}}\|} \tag{15b}$$

The atmospheric density $\rho$ is calculated using the exponential atmosphere model $\rho = \rho_0 e^{-(h)/h_s}$, where $\rho_0 = 1.225$ kg/m$^3$ is the density at sea level, and $h_o = 7018.00344$ m is the density scale-height, and the Mach number $M$ is calculated using the standard Earth atmospheric temperature model. The missile control surface deflections $\boldsymbol{\theta}_{\text{ctrl}} = [\theta_{\text{LT}}, \theta_{\text{RT}}, \theta_{\text{BT}}, \theta_{\text{UT}}]$, dynamic pressure $q_o = \frac{1}{2}\rho V^2$ with $V = \|\mathbf{v}^B\|$, rotational velocity vector $\boldsymbol{\omega}$, $M$, $\alpha$ and $\beta$ are then input to the aerodynamic model described in Section II.F to compute the body frame force $\mathbf{F}^{\text{B}}$ and torque $\mathbf{L}^{\text{B}}$.

The rotational velocity $\boldsymbol{\omega}$ is updated by integrating the Euler rotational equations of motion, as shown in Equation (16), where $\mathbf{J}$ is the missile's inertia tensor, and with the skew symmetric operator $[\mathbf{a}\times]$ defined in Eq. (17).

$$\mathbf{J}\dot{\boldsymbol{\omega}} = -[\boldsymbol{\omega}\times]\mathbf{J}\boldsymbol{\omega} + \mathbf{L}^{\text{B}} \tag{16}$$

$$[\mathbf{a}\times] \equiv \begin{bmatrix} 0 & -a_3 & a_2 \\ a_3 & 0 & -a_1 \\ -a_2 & a_1 & 0 \end{bmatrix} \tag{17}$$

The missile's attitude $\mathbf{q}$ is updated by integrating the differential kinematic equations shown in Equation (18), where the missile's attitude is parameterized using the quaternion representation and $\omega_i$ denotes the $i^{th}$ component of the rotational velocity vector $\boldsymbol{\omega}_B$.

$$\begin{bmatrix} \dot{q}_0 \\ \dot{q}_1 \\ \dot{q}_2 \\ \dot{q}_3 \end{bmatrix} = \frac{1}{2} \begin{bmatrix} q_0 & -q_1 & -q_2 & -q_3 \\ q_1 & q_0 & -q_3 & q_2 \\ q_2 & q_3 & q_0 & -q_1 \\ q_3 & -q_2 & q_1 & q_0 \end{bmatrix} \begin{bmatrix} 0 \\ \omega_0 \\ \omega_1 \\ \omega_2 \end{bmatrix} \tag{18}$$

The missile body frame velocity $\mathbf{v}^{\mathrm{B}}$ is updated by integrating the differential equation given in Eq. 19, where $m$ is the missile mass, $\mathbf{C}_{\mathrm{BN}}(\mathbf{q})$ is the DCM mapping from the inertial to body frame given the missile attitude $\mathbf{q}$, and $g = -9.81$ m/s$^2$.

$$\dot{\mathbf{v}}^{\mathrm{B}} = -[\omega\times]\mathbf{v}^{\mathrm{B}} + \frac{\mathbf{F}^{\mathrm{B}}}{m} + [\mathbf{C}_{\mathrm{BN}}(\mathbf{q})]^T \begin{bmatrix} 0 & 0 & g \end{bmatrix} \tag{19}$$

$\mathbf{v}^{\mathrm{B}}$ is then rotated into the North-East-Down (NED) inertial frame as shown in Eq. (20a), and $\mathbf{v}_{\mathrm{M}}$ is calculated by rotating $\mathbf{v}^{\mathrm{NED}}$ 180° around the x-axis into the reference frame depicted in Fig. 2.

$$\mathbf{v}^{\mathrm{NED}} = \mathbf{C}_{\mathrm{BN}}(\mathbf{q})\mathbf{v}^{\mathrm{B}} \tag{20a}$$

$$\mathbf{v}_{\mathrm{M}} = \begin{bmatrix} \mathbf{v}_0^{\mathrm{NED}} & -\mathbf{v}_1^{\mathrm{NED}} & -\mathbf{v}_2^{\mathrm{NED}} \end{bmatrix} \tag{20b}$$

The missile inertial frame position $\mathbf{r}_{\mathrm{M}}$ is then updated by integrating Eq. (21)

$$\dot{\mathbf{r}}_{\mathrm{M}} = \mathbf{v}_{\mathrm{M}} \tag{21}$$

The target is modeled as shown in Eqs. (22a) through (22b), where $\mathbf{a}_{\mathrm{T}}$ is the target acceleration assuming the maneuvers described in Section II.A

$$\dot{\mathbf{r}}_{\mathrm{T}} = \mathbf{v}_{\mathrm{T}} \tag{22a}$$

$$\dot{\mathbf{v}}_{\mathrm{T}} = \mathbf{a}_{\mathrm{T}} \tag{22b}$$

The equations of motion are updated using fourth order Runge-Kutta integration. For ranges greater than 40 m, a timestep of 10 ms is used, and for the final 40 m of homing, a timestep of 0.1 ms is used in order to more accurately (within 0.2m) measure miss distance; this technique is borrowed from [25].

# III. Methods

## A. Reinforcement Learning Framework

In the reinforcement learning framework, an agent learns through episodic interaction with an environment how to successfully complete a task using a policy that maps observations $\mathbf{o}$ to actions $\mathbf{u}$. The environment initializes an episode by randomly generating a ground truth state $\mathbf{x}$, mapping this state to an observation, and passing the observation to the agent. The agent uses this observation to generate an action that is sent to the environment; the environment then uses the action and the current ground truth state to generate the next state and a scalar reward signal $r(\mathbf{x}, \mathbf{u})$. The reward and the observation corresponding to the next state are then passed to the agent. The process repeats until the environment terminates the episode, with the termination signaled to the agent via a done signal. Trajectories collected over a set of episodes (referred to as rollouts) are collected during interaction between the agent and environment, and used to update the policy and value functions. The interface between agent and environment is depicted in Fig. 10, where the environment instantiates the models shown in Fig. 1

Meta-RL differs from generic reinforcement learning in that the agent learns over an ensemble of environments. These environments vary the engagement scenarios, dynamics, aerodynamic coefficients, radome parameters, and other factors. Optimization within the meta-RL framework results in an agent that can quickly adapt to novel environments, often with just a few steps of interaction with the environment. Similar to [26], we implement meta-RL by including a recurrent layer in the policy and value function. For a given trajectory over observations and actions, the recurrent layer will evolve differently in environments instantiating different aerodynamic models. By optimizing over an ensemble of these aerodynamic models, the agent learns to adapt, using the recurrent layer's hidden state to infer the current aerodynamic model.

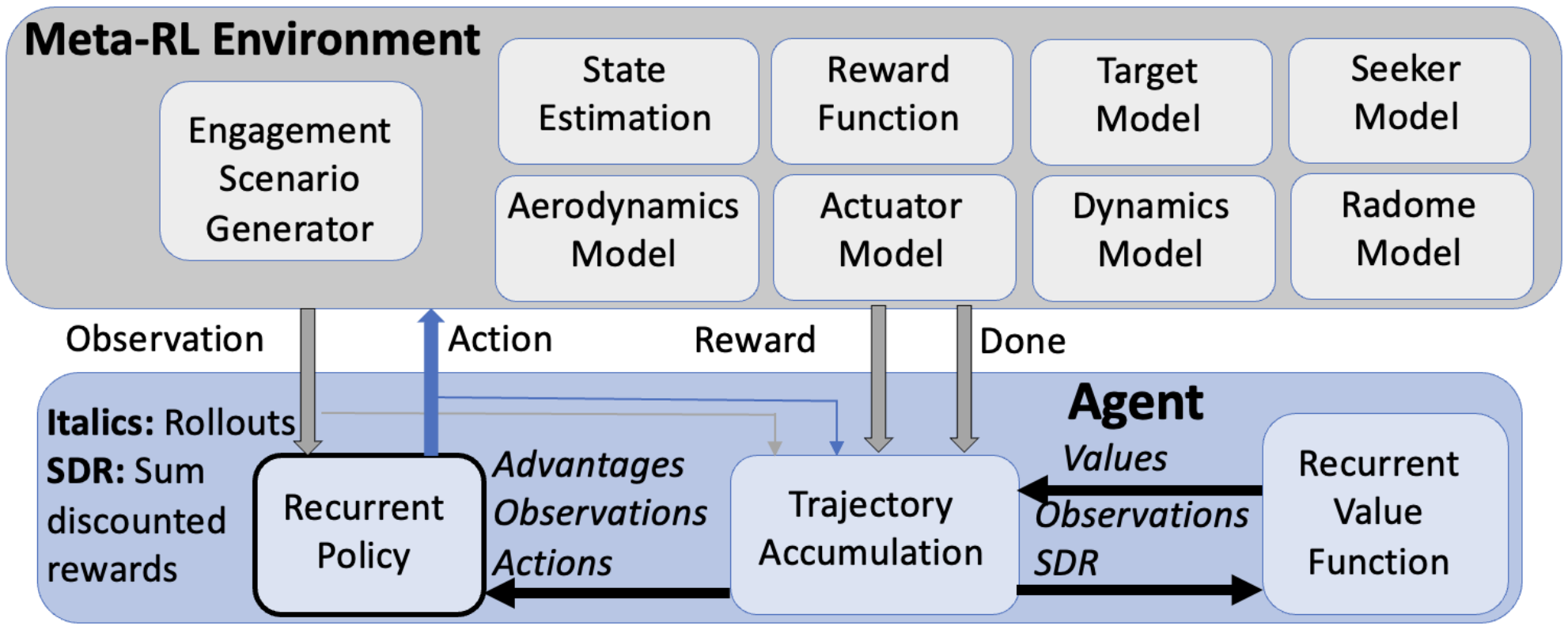

**Fig. 10 Environment-Agent Interface**

The PPO algorithm used in this work is a policy gradient algorithm which has demonstrated state-of-the-art performance for many reinforcement learning benchmark problems. PPO approximates the Trust Region Policy Optimization method [27] by accounting for the policy adjustment constraint with a clipped objective function. The objective function used with PPO can be expressed in terms of the probability ratio $p_k(\boldsymbol{\theta})$ given by,

$$p_k(\boldsymbol{\theta}) = \frac{\pi_{\theta}(\mathbf{u}_k|\mathbf{o}_k)}{\pi_{\theta_{\text{old}}}(\mathbf{u}_k|\mathbf{o}_k)} \tag{23}$$

The PPO objective function is shown in Equations (24a) through (24c). The general idea is to create two surrogate objectives, the first being the probability ratio $p_k(\boldsymbol{\theta})$ multiplied by the advantages $A^{\pi}_{\mathbf{w}}(\mathbf{o}_k, \mathbf{u}_k)$ (see Eq. (25)), and the second a clipped (using clipping parameter $\epsilon$) version of $p_k(\boldsymbol{\theta})$ multiplied by $A^{\pi}_{\mathbf{w}}(\mathbf{o}_k, \mathbf{u}_k)$. The objective to be maximized $J(\boldsymbol{\theta})$ is then the expectation under the trajectories induced by the policy of the lesser of these two surrogate objectives.

$$\text{obj1} = p_k(\boldsymbol{\theta})A^{\pi}_{\mathbf{w}}(\mathbf{o}_k, \mathbf{u}_k) \tag{24a}$$

$$\text{obj2} = \text{clip}(p_k(\boldsymbol{\theta})A^{\pi}_{\mathbf{w}}(\mathbf{o}_k, \mathbf{u}_k), 1-\epsilon, 1+\epsilon) \tag{24b}$$

$$J(\boldsymbol{\theta}) = \mathbb{E}_{p(\tau)}[\min(\text{obj1}, \text{obj2})] \tag{24c}$$

This clipped objective function has been shown to maintain a bounded Kullback-Leibler (KL) divergence [28] with respect to the policy distributions between updates, which aids convergence by ensuring that the policy does not change drastically between updates. Our implementation of PPO uses an approximation to the advantage function that is the difference between the empirical return and a state value function baseline, as shown in Equation 25, where $\gamma$ is a discount rate and $r$ the reward function, described in Section III.B.

$$A^{\pi}_{\mathbf{w}}(\mathbf{x}_k, \mathbf{u}_k) = \left[\sum_{\ell=k}^{T} \gamma^{\ell-k} r(\mathbf{x}_\ell, \mathbf{u}_\ell)\right] - V^{\pi}_{\mathbf{w}}(\mathbf{x}_k) \tag{25}$$

Here the value function $V^{\pi}_{\mathbf{w}}$ is learned using the cost function given by

$$L(\mathbf{w}) = \frac{1}{2M}\sum_{i=1}^{M}\left(V^{\pi}_{\mathbf{w}}(\mathbf{x}^i_k) - \left[\sum_{\ell=k}^{T} \gamma^{\ell-k} r(\mathbf{u}^i_\ell, \mathbf{x}^i_\ell)\right]\right)^2 \tag{26}$$

In practice, policy gradient algorithms update the policy using a batch of trajectories (roll-outs) collected by interaction with the environment. Each trajectory is associated with a single episode, with a sample from a trajectory collected at step $k$ consisting of observation $\mathbf{o}_k$, action $\mathbf{u}_k$, and reward $r_k(\mathbf{o}_k, \mathbf{u}_k)$. Finally, gradient ascent is performed on $\boldsymbol{\theta}$ and gradient descent on $\mathbf{w}$ and update equations are given by

$$\mathbf{w}^{+} = \mathbf{w}^{-} - \beta_{\mathbf{w}} \nabla_{\mathbf{w}} \, L(\mathbf{w})|_{\mathbf{w}=\mathbf{w}^{-}} \tag{27}$$

$$\boldsymbol{\theta}^{+} = \boldsymbol{\theta}^{-} + \beta_{\boldsymbol{\theta}} \, \nabla_{\boldsymbol{\theta}} J\,(\boldsymbol{\theta})|_{\boldsymbol{\theta}=\boldsymbol{\theta}^{-}} \tag{28}$$

where $\beta_{\mathbf{w}}$ and $\beta_{\boldsymbol{\theta}}$ are the learning rates for the value function, $V_{\mathbf{w}}^{\pi}(\mathbf{o}_k)$, and policy, $\pi_{\boldsymbol{\theta}}(\mathbf{u}_k|\mathbf{o}_k)$, respectively.

In our implementation of PPO, we adaptively scale the observations and servo both $\epsilon$ and the learning rate to target a KL divergence of 0.001.

### B. Meta-RL Problem Formulation

In this air-to-air missile application, an episode terminates when the closing velocity $v_c$ turns negative, the missile speed falls below 400 m/s, or a path constraint is violated (See Table 2). The agent observation $\mathbf{o}$ is shown in Eq. (29), where $\boldsymbol{\lambda}^{\mathrm{S}}_{\mathrm{obs}}$, $\boldsymbol{\Omega}$, $r_{\mathrm{obs}}$, $v_{c_{\mathrm{obs}}}$, $\mathbf{q}_{\mathrm{obs}}$, $\mathbf{a}^{\mathrm{B}}_{\mathrm{obs}}$, and $\boldsymbol{\omega}^{\mathrm{B}}_{\mathrm{obs}}$ are generated in the seeker model (Section II.B), LOS stabilization (Section II.C), and filtering and state estimation block (Section II.D), and $\boldsymbol{\theta}_{\mathrm{defl}}$ are the commanded deflections (the input to the actuator model) from Section II.E. The policy actions are interpreted as described in Section II.E.

$$\mathbf{o} = \begin{bmatrix} \boldsymbol{\lambda}^{\mathrm{S}}_{\mathrm{obs}} & \boldsymbol{\Omega} & v_{c_{\mathrm{obs}}} & r_{\mathrm{obs}} & \mathbf{q}_{\mathrm{obs}} & \boldsymbol{\omega}^{\mathrm{B}}_{\mathrm{obs}} & \mathbf{a}^{\mathrm{B}}_{\mathrm{obs}} & \boldsymbol{\theta}_{\mathrm{defl}} \end{bmatrix} \tag{29}$$

The reward function is shown below in Equations (30a) through (30f). $r_{\mathrm{shaping}}$ is a shaping reward given at each step in an episode. These shaping rewards take the form of a Gaussian-like function of the norm of the line of sight rotation rate $\boldsymbol{\Omega}$. $r_{\mathrm{rollrate}}$ encourages the agent to minimize roll rate, with differential fin deflections only used to counteract the rolling moment induced by angle of attack and sideslip (See Section II.F). $r_{\mathrm{ctrl}}$ is a control effort penalty, again given at each step in an episode, and $r_{\mathrm{bonus}}$ is a bonus given at the end of an episode if the miss distance is below the threshold $r_{\mathrm{lim}}$. Importantly, the current episode is terminated if a path constraint is violated, in which case the stream of positive shaping rewards is terminated, and the agent receives a negative reward. We use $\alpha = 1$, $\beta = -0.05$, $\delta = -0.01$, $\epsilon = 10$, $\zeta = -10$, $r_{\mathrm{lim}} = 10$ m, $\sigma_{\Omega} = 0.02$. We use a discount rate of 0.95 for shaping rewards and 0.995 for the terminal reward.

$$r_{\mathrm{shaping}} = \alpha \exp\left(\frac{-\,\|\boldsymbol{\Omega}\|^2}{\sigma_{\Omega}^2}\right) \tag{30a}$$

$$r_{\mathrm{rollrate}} = \beta|\omega_x| \tag{30b}$$

$$r_{\mathrm{ctrl}} = \delta\left\|\begin{bmatrix}\theta_{\mathrm{LT}} & \theta_{\mathrm{RT}} & \theta_{\mathrm{BT}} & \theta_{\mathrm{UT}}\end{bmatrix}\right\| \tag{30c}$$

$$r_{\mathrm{bonus}} = \begin{cases} \epsilon, & \text{if } \mathbf{r}_{\mathrm{TM}} < r_{\mathrm{lim}} \text{ and done} \\ 0, & \text{otherwise} \end{cases} \tag{30d}$$

$$r_{\mathrm{penalty}} = \begin{cases} \zeta, & \text{if any path constraint violated} \\ 0, & \text{otherwise} \end{cases} \tag{30e}$$

$$r = r_{\mathrm{shaping}} + r_{\mathrm{ctrl}} + r_{\mathrm{bonus}} + r_{\mathrm{penalty}} \tag{30f}$$

The policy and value functions are implemented using four layer neural networks with tanh activations on each hidden layer. Layer 2 for the policy and value function is a recurrent layer implemented using gated recurrent units [29]. The network architectures are as shown in Table 5, where $n_{\mathrm{h}i}$ is the number of units in layer $i$, obs_dim is the observation dimension, and act_dim is the action dimension. The policy and value functions are periodically updated during optimization after accumulating trajectory rollouts of 30 simulated episodes.

**Table 5 Policy and Value Function network architecture**

| | Policy Network | | Value Network | |
|---|---|---|---|---|
| Layer | # units | activation | # units | activation |
| hidden 1 | $10*\text{obs_dim}$ | tanh | $10*\text{obs_dim}$ | tanh |
| hidden 2 | $\sqrt{n_{h1} * n_{h3}}$ | tanh | $\sqrt{n_{h1} * n_{h3}}$ | tanh |
| hidden 3 | $10 * \text{act_dim}$ | tanh | 5 | tanh |
| output | act_dim | linear | 1 | linear |

# IV. Experiments

## A. Longitudinal Three Loop Autopilot Benchmark

We evaluated the longitudinal autopilot model from Zarchan [30] using the adjoint analysis code in Listing 23.3, but modified the code to run a range of altitudes from 0 to 15km[†] in 1.5 km steps, and modified some of the parameters for better comparison with our results. Specifically, we increased the actuator bandwidth to 600 rad/s, limited the radome slope to between -0.01 and 0.01, removed glint and fading noise, and reduced range independent noise amplitude from 2 mrad to 1 mrad. Other parameters were tuned to increase performance over the reduced range of radome slope. In particular, we decreased the flight control system time constant to 0.10, the flight control system damping to 0.4, the noise filter time constant to 0.1s, and increased the guidance navigation ratio to 4. This maximized performance over the simulated altitude range, with further reduction of the seeker or noise filter time constants being counterproductive at higher altitudes. Similarly, attempting to increase the guidance navigation ratio above 4 led to catastrophic instability. The adjoint simulation used a target step maneuver and did not model drag, which kept the closing speed constant at 1250 m/s. Note that we scaled the target acceleration taking into account dynamic pressure using the method described in Section II.A. Importantly, the flight control system gains are recomputed at each step using the ground truth missile altitude and speed. The results are given in Table 6 for various altitudes and target acceleration levels. The miss statistics are computed over the root-mean-square miss taking into account target maneuver, noise, and radome slope, with radome slopes varying from -0.01 to 0.01 in increments of 0.001 for each altitude, We see that the benchmark system is accurate, but accuracy within the 100cm limit falls off quickly with increased target acceleration.

**Table 6 Longitudinal PN + Three Loop Autopilot Benchmark**

| TACC (g) | $\mu$ Miss (m) | $\sigma$ Miss (m) | Miss < 100cm (%) | Miss < 200cm (%) | Miss < 300cm (%) |
|---|---|---|---|---|---|
| 2 | 0.7 | 0.3 | 95 | 99 | 100 |
| 4 | 0.9 | 0.3 | 71 | 99 | 100 |
| 6 | 1.1 | 0.4 | 42 | 99 | 100 |
| 8 | 1.4 | 0.5 | 25 | 81 | 100 |

## B. Meta-RL Optimization and Testing

Optimization uses the initial conditions and vehicle parameters given in the problem formulation (Section II). Aside from rare violations due to agent exploration, the agent quickly learns to satisfy the path constraints. Once the agent learns to satisfy the constraints, the agent adjusts its policy to maximize both shaping and terminal rewards while continuing to satisfy constraints. Learning curves are given in Figures 11 through 12. Fig. 11 plots the mean, mean - 1 standard deviation, and minimum rewards plotted on the primary y-axis and the mean and maximum number of steps per episode plotted on the secondary y-axis. Similarly, Fig. 12 plots terminal miss distance statistics. These statistics are computed over a batch of rollouts (30 episodes). The occasional large miss distances are due to exploration, which is why we turn exploration off for the deployed policy.

To test the meta-RL G&C system, we ran 5000 episodes over each of the cases shown in Table 7, where $\|\mathbf{a}_M\|$ is the norm of the body frame normal and side accelerations. The "Nominal" case is identical to that used for optimization, the "High Altitude" case has the missile starting at an altitude of 15km for all engagements, the "TACC=$X$" cases have maximum target acceleration set to $X$, the "Peak $\theta_R = X$" cases have $A_u$ and $A_v$ increased to $X$ in the radome model (see Section II.B), and the "PV=$X$" cases have the maximum aerodynamic coefficient variation set to $X$ ($\epsilon_{\text{coeff}}$ in Table 4). The last three entries in Table 7 ("STEP-TACC=X") use a step target maneuver with the target acceleration set to "X"; these statistics are computed using the same initial conditions as in the nominal case, using the full 6-DOF model, but with the initial missile and target speed set to 1000 m/s and 300 m/s respectively. This gives a more direct comparison to the benchmark results from Section IV.A. However, as compared to the longitudinal benchmark, the missile using the meta-RL G&C system has a much more difficult problem, as the aerodynamics take into account drag, roll coupling effects, aerodynamic parameter variation, and the initial conditions cover a range of initial heading errors, angles of attack, sideslip angles, and roll angles.

[†] Although Imperial units are used in these simulations, we convert them to metric in the discussion for consistency

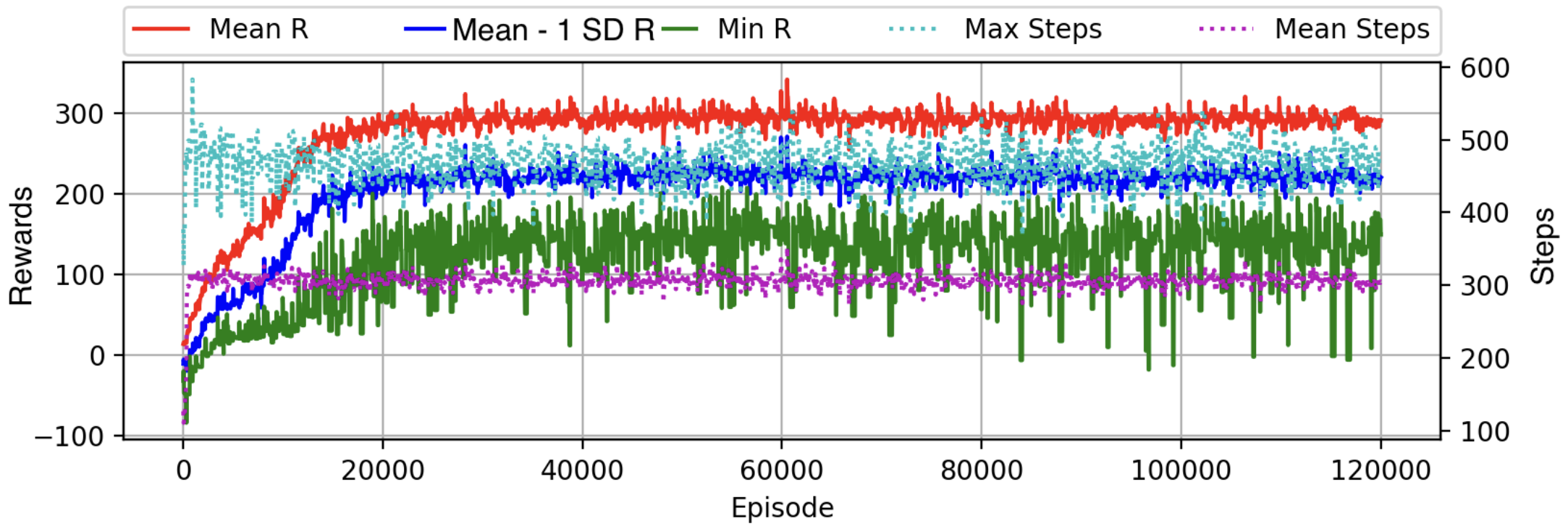

**Fig. 11 Optimization Reward History**

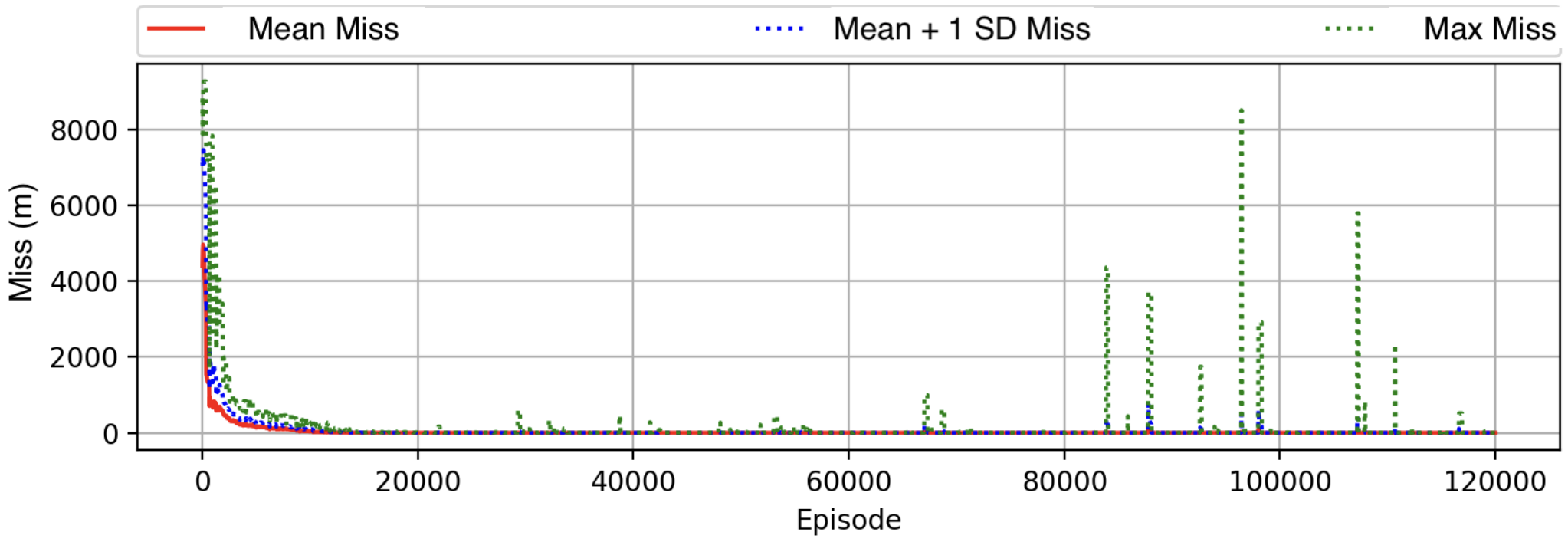

**Fig. 12 Optimization Miss Distance History**

The statistics for the target acceleration for the "Nominal" case simulations are a mean, standard deviation, and maximum of 5.9m/s$^2$, 7.5m/s$^2$,and 67.2m/s$^2$, respectively. A trajectory starting at an altitude of 15 km that resulted in a 0.7 m miss is shown in Fig. 13. Comparing this to the open loop response shown in Fig. 9, we see that the adaptive G&C system implements an effective guidance and flight control system. Note that the target acceleration (shown in the inertial frame) is adjusted to account for dynamic pressure as described in Section II.A and is less than the maximum acceleration given in Table 7. Missile acceleration is shown in the body frame, with "LY2D" being the norm of the lift and side forces to drag force.

### C. Performance with Flexible Body Dynamics

We test the system optimized in Section IV.B using a flexible body dynamics model that employs an approach similar to that described in [31]. The equations governing the flexible body dynamics are shown in Eqs. (31a) through (31d), where the superscript $i$ indicates the $i^{\text{th}}$ flexible mode, $F_2^{\text{B}}$ is the body frame normal force , and $F_1^{\text{B}}$ is the body frame side force, as computed in Section II.F Eq. 13a. Note that the flexible body dynamics are in addition to the aerodynamic coefficient variation shown in Table 4.

**Table 7 Meta-RL Policy Performance**

| | Miss < 1m | Miss < 2m | Miss < 3m | $V_f$ (m/s) | | $\|\mathbf{a}_{\mathrm{M}}\|$ (m/s$^2$) | | Vio |
|---|---|---|---|---|---|---|---|---|
| Case | % | % | % | $\mu$ | $\sigma$ | $\mu$ | $\sigma$ | % |
| Nominal | 91 | 98 | 99 | 805 | 64 | 21 | 29 | 0.0 |
| High Altitude | 81 | 87 | 99 | 833 | 60 | 16 | 19 | 0.0 |
| TACC=10 g | 89 | 97 | 99 | 804 | 65 | 22 | 30 | 0.0 |
| TACC=12 g | 86 | 96 | 99 | 800 | 66 | 24 | 31 | 0.7 |
| TACC=15 g | 82 | 94 | 97 | 796 | 70 | 25 | 32 | 1.0 |
| TACC=20 g | 76 | 89 | 93 | 786 | 78 | 29 | 35 | 1.4 |
| Peak $\theta_R$ = 0.02 | 77 | 95 | 98 | 804 | 63 | 21 | 30 | 0.6 |
| Peak $\theta_R$ = 0.05 | 46 | 72 | 84 | 801 | 64 | 23 | 32 | 1.8 |
| PV=20% | 88 | 97 | 98 | 804 | 65 | 22 | 31 | 1.3 |
| PV=30% | 79 | 90 | 93 | 795 | 77 | 25 | 36 | 3.4 |
| STEP-TACC=2 g | 98 | 99 | 99 | 895 | 43 | 21 | 32 | 0.0 |
| STEP-TACC=4 g | 98 | 99 | 99 | 887 | 48 | 27 | 31 | 0.0 |
| STEP-TACC=6 g | 96 | 98 | 98 | 874 | 60 | 33 | 33 | 0.2 |
| STEP-TACC=8 g | 92 | 96 | 97 | 862 | 70 | 39 | 35 | 1.1 |

$$\ddot{\eta}_{\mathrm{N}}^{(1)} = -2\zeta_{\mathrm{fb}}^{(1)}\omega_{\mathrm{fb}}^{(1)}\dot{\eta}_{\mathrm{N}}^{(1)} - \left(\omega_{\mathrm{fb}}^{(1)}\right)^2\eta_{\mathrm{N}}^{(1)} + F_2^{\mathrm{B}} \tag{31a}$$

$$\ddot{\eta}_{\mathrm{N}}^{(2)} = -2\zeta_{\mathrm{fb}}^{(2)}\omega_{\mathrm{fb}}^{(2)}\dot{\eta}_{\mathrm{N}}^{(2)} - \left(\omega_{\mathrm{fb}}^{(2)}\right)^2\eta_{\mathrm{N}}^{(2)} + F_2^{\mathrm{B}} \tag{31b}$$

$$\ddot{\eta}_{\mathrm{Y}}^{(1)} = -2\zeta_{\mathrm{fb}}^{(1)}\omega_{\mathrm{fb}}^{(1)}\dot{\eta}_{\mathrm{Y}}^{(1)} - \left(\omega_{\mathrm{fb}}^{(1)}\right)^2\eta_{\mathrm{Y}}^{(1)} + F_1^{\mathrm{B}} \tag{31c}$$

$$\ddot{\eta}_{\mathrm{Y}}^{(2)} = -2\zeta_{\mathrm{fb}}^{(2)}\omega_{\mathrm{fb}}^{(2)}\dot{\eta}_{\mathrm{Y}}^{(2)} - \left(\omega_{\mathrm{fb}}^{(2)}\right)^2\eta_{\mathrm{Y}}^{(2)} + F_1^{\mathrm{B}} \tag{31d}$$

The normal and side force coefficients (Section II.F) are then perturbed as shown in Eqs. (32a) through (32b). Importantly, the perturbation to the pitching and yawing moment coefficients couples into the torque (Section II.F), which perturbs the measured rotational velocity vector $\boldsymbol{\omega}_{\mathrm{obs}}^{\mathrm{B}}$, creating a parasitic feedback path between control surface deflections and $\boldsymbol{\omega}_{\mathrm{obs}}^{\mathrm{B}}$ [32]. This parasitic feedback path impacts LOS stabilization, as described in Section II.C.

$$\mathbf{C}_{N_{\mathrm{fb}}} = \mathbf{C}_N\left(1 + k^{(1)}\eta_{\mathrm{N}}^{(1)} + k^{(2)}\eta_{\mathrm{N}}^{(2)}\right) \tag{32a}$$

$$\mathbf{C}_{Y_{\mathrm{fb}}} = \mathbf{C}_Y\left(1 + k^{(1)}\eta_{\mathrm{Y}}^{(1)} + k^{(2)}\eta_{\mathrm{Y}}^{(2)}\right) \tag{32b}$$

$$\mathbf{C}_{m_{\mathrm{fb}}} = \mathbf{C}_m\left(1 + k^{(1)}\eta_{\mathrm{N}}^{(1)} + k^{(2)}\eta_{\mathrm{N}}^{(2)}\right) \tag{32c}$$

$$\mathbf{C}_{n_{\mathrm{fb}}} = \mathbf{C}_n\left(1 + k^{(1)}\eta_{\mathrm{Y}}^{(1)} + k^{(2)}\eta_{\mathrm{Y}}^{(2)}\right) \tag{32d}$$

We use $\omega_{\mathrm{fb}}^{(1)} = 169$ rad/s, $\omega_{\mathrm{fb}}^{(1)} = 400$ rad/s, $\zeta_{\mathrm{fb}}^{(1)} = 0.015$, and $\zeta_{\mathrm{fb}}^{(2)} = 0.022$. The reason we tested, but did not optimize a system with consideration of flexible body dynamics is that the high frequency flexible modes require simulating with an integration time step of 1 ms, which slows down the simulations by a factor of 10. To give an idea of the magnitude of the aerodynamic force perturbations, we captured statistics for the perturbation of $\mathbf{C}_N$ and $\mathbf{C}_Y$, with the perturbations calculated as the absolute value of the percentage change between the aerodynamic coefficient perturbed as shown in Table 4 and the additional perturbation given in Eqs. (32a) and (32b). The simulation results are tabulated in Table 8. We see that with the exception of $k^{(0)} = k^{(1)} = 0.1$, the results are similar to the "Nominal" case in Table 7. Interestingly, the maximum perturbations have values that are over 20 times the standard deviation of the perturbation distribution. For the last two cases in Table 8, these result in constraint violations. It is possible that if we had optimized with flexible body dynamics, the agent would have learned to avoid sequences of actions that lead to these excessive perturbations.

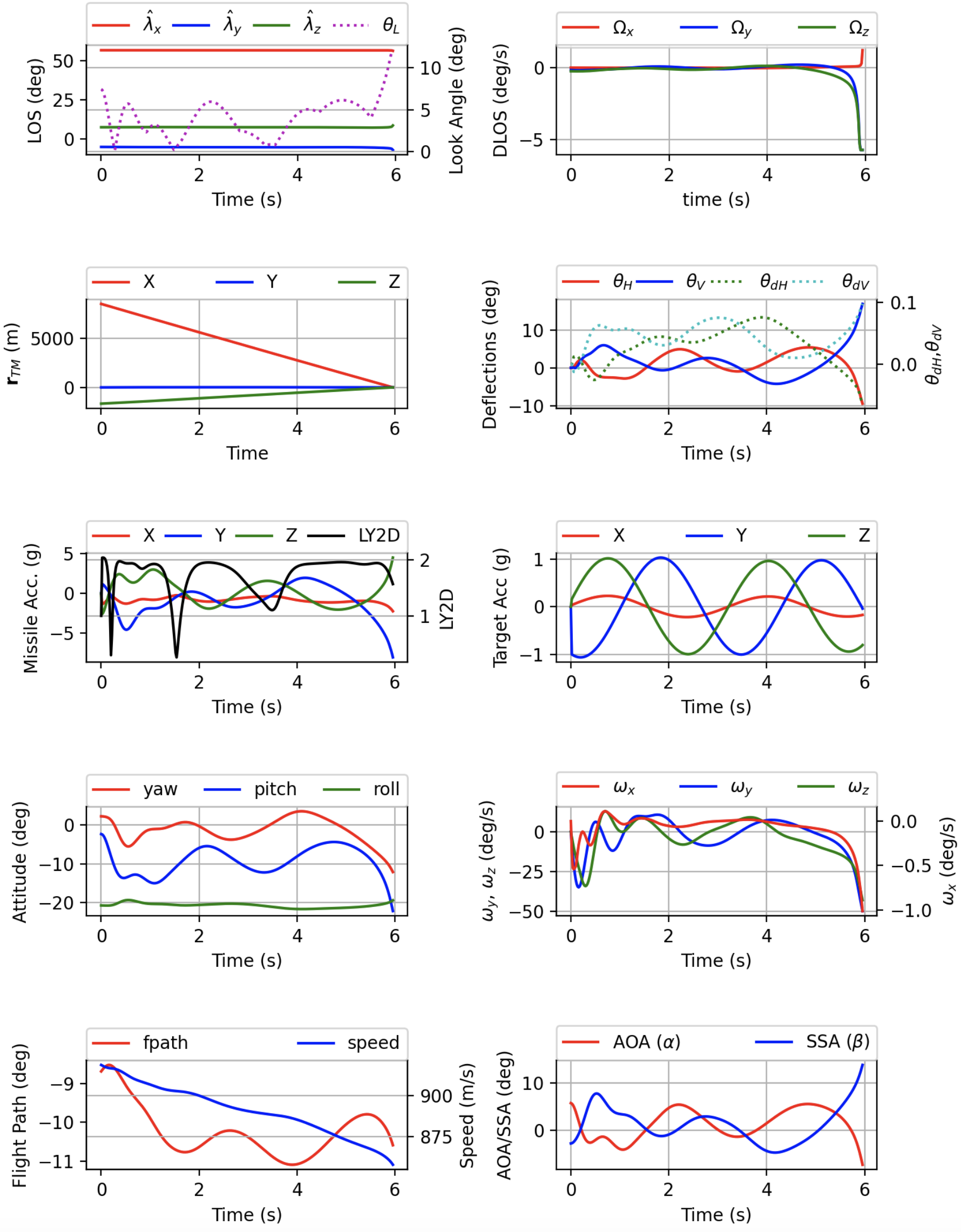

**Fig. 13 High Altitude Sample Trajectory from Adaptive System**

**Table 8 Performance with Flexible Body Dynamics**

| Case | $\Delta_N$ (%) | | | $\Delta_Y$ (%) | | | Miss < 1m | Miss < 2m | Miss < 3m | Vio |
|---|---|---|---|---|---|---|---|---|---|---|
| $k^{(0)} = k^{(1)}$ | $\mu$ | $\sigma$ | Max | $\mu$ | $\sigma$ | Max | % | % | % | % |
| 0.02 | 0.7 | 0.9 | 30.0 | 0.6 | 0.9 | 19.4 | 91 | 99 | 100 | 0.2 |
| 0.05 | 1.8 | 2.4 | 69.9 | 1.4 | 2.3 | 48.6 | 91 | 98 | 99 | 1.1 |
| 0.10 | 3.7 | 4.9 | 133.5 | 2.8 | 4.7 | 97.7 | 88 | 94 | 95 | 4.5 |

### D. Performance without Range and Range Rate

Missiles using infrared seekers do not have access to range and range rate measurements. Consequently, the exact LOS rotation rate $\boldsymbol{\Omega} = \frac{\mathbf{r}^{S}_{TM} \times \mathbf{v}^{S}_{TM}}{\mathbf{r}^{S}_{TM} \cdot \mathbf{r}^{S}_{TM}}$ cannot be calculated. However, we can optimize a system using the surrogate LOS rotation rate described in Section II.D, and calculated as $\boldsymbol{\Omega} = \frac{\hat{\boldsymbol{\lambda}}_t \times \hat{\boldsymbol{\lambda}}_{t-\Delta t}}{\Delta t}$. We optimized such a system, with $r$ and $v_c$ removed from the observation vector given in Eq. 29. Performance is shown in Table 9. Performance was reduced as compared to that shown in Section IV.B, but still exceeds that of the benchmark, and could likely be improved upon with further hyperparameter tuning. One issue we found using the surrogate LOS rotation rate is reduced tolerance to seeker noise, and we had to reduce the noise standard deviation used in Section IV.B (1 mrad/s) to 0.5 mrad/s to obtain the performance reported in Table 9.

**Table 9 Meta-RL Policy Performance without Range and Range Rate**

| | Miss < 1m | Miss < 2m | Miss < 3m | $V_f$ (m/s) | | $\|\|\mathbf{a}_M\|\|$ (m/s$^2$) | | Vio |
|---|---|---|---|---|---|---|---|---|
| Case | % | % | % | $\mu$ | $\sigma$ | $\mu$ | $\sigma$ | % |
| Nominal | 88 | 97 | 98 | 815 | 63 | 18 | 21 | 0.0 |
| TACC=10 g | 81 | 94 | 97 | 811 | 65 | 18 | 23 | 0.3 |
| TACC=12 g | 78 | 92 | 96 | 811 | 67 | 21 | 26 | 0.6 |
| TACC=15 g | 71 | 86 | 93 | 803 | 73 | 23 | 29 | 0.9 |
| TACC=20 g | 64 | 78 | 86 | 791 | 81 | 27 | 34 | 1.6 |
| PV=20% | 85 | 96 | 98 | 813 | 66 | 18 | 22 | 0.3 |
| PV=30% | 80 | 92 | 95 | 807 | 73 | 19 | 25 | 1.0 |
| High Altitude | 81 | 93 | 96 | 842 | 62 | 15 | 13 | 0.0 |
| STEP-TACC=2 × 9.81 | 100 | 100 | 100 | 812 | 63 | 16 | 16 | 0.0 |
| STEP-TACC=4 × 9.81 | 99 | 100 | 100 | 803 | 68 | 22 | 16 | 0.0 |
| STEP-TACC=6 × 9.81 | 98 | 99 | 100 | 790 | 74 | 29 | 18 | 0.0 |
| STEP-TACC=8 × 9.81 | 93 | 97 | 98 | 771 | 82 | 35 | 20 | 0.5 |

### E. Performance without Recurrent Network Layers

Here we optimized a G&C policy using the same framework, but without the recurrent layers in the policy and value function networks. Although this can result in a robust system, the system is not adaptive, in that without the recurrent layers its behavior is fixed at deployment. When tested against the "Nominal" case the G&C system managed to achieve a miss distances of 1m, 2m, and 3m in 10%, 17%, and 22% of simulated engagement, and many trajectories had miss distances of several km. In order to test whether the poor performance was due to the inability to measure altitude and speed, we repeated the experiment, but added altitude and speed to the observation. We found that performance did not improve, suggesting that an effective guidance and flight control system must map a history of inputs to commanded deflection rates, as opposed to mapping the current observation to commanded deflection rates. This should come as no surprise, as a three loop autopilot integrates the difference between the acceleration tracking error and the rotational velocity measured for that channel [30], with the integrator state being a function of the history of this difference. Thus, it appears that a single loop integrated G&C system implemented as a neural network must contain at least one recurrent layer in the policy.

### F. Discussion

Without measurements of altitude or speed, the meta-RL optimized G&C system adapts well to the large flight envelope (0 to 20km altitude) used in our experiments, where the missile dynamic pressure ranges from 22,680 kg-m$^2$ at 20 km and 800 m/s to 612,500 kg-m$^2$ at sea level and 1000 m/s. From Table 7 we see that the G&C system adapts well to novel conditions not experienced during optimization. Although performance decreased with higher peak radome aberration angles, this could be improved using an active compensation approach [33], and it is possible that optimizing with higher radome slopes may improve tolerance to larger aberration angles. Importantly, the policy adapts well to variation in aerodynamic force and moment coefficients, as well as flexible body dynamics that were not modeled during optimization, demonstrating that the system is robust to differences between the optimization and deployment environments. We see that performance deteriorates with higher target acceleration levels, and this could likely be improved by incorporating bias into the reward shaping function, similar to biased PN [34]. Alternately, a meta-RL optimized LOS shaping policy [35] has been demonstrated to improve accuracy against high acceleration target maneuvers. G&C for systems without range or range rate measurements (as is the case with infrared seekers) is a challenging problem. Nevertheless, although the G&C system optimized without range or range rate (Section IV.D) had reduced performance as compared to the system optimized in Section IV.B, performance still exceeded that of the benchmark system.

The meta-RL optimized G&C system outperformed the longitudinal benchmark of PN coupled with a three loop autopilot by a large margin. This is despite the fact that the benchmark used linearized airframe and dynamics models, did not model drag, roll coupling, actuator saturation, or aerodynamic parameter variation, used a less challenging target maneuver, and computed flight control system gains using the ground truth missile altitude and speed. Further, the benchmark was a simpler control problem, with three rather than six degrees of freedom, and did not address roll stabilization. We believe that the meta-RL policy outperformed the benchmark due to a combination of three factors. First, the meta-RL policy takes advantage of the underdamped airframe, particularly early in the engagement. We see from Figs. 8 and 9 that the open loop airframe response is underdamped, with the achieved acceleration overshooting the steady state acceleration. Clearly, a G&C system that can take advantage of this behavior can obtain a performance advantage. In contrast, in the traditional approach where a guidance system is coupled with a flight control system, there is a more limited scope for decreasing the flight control system damping, as this can lead to decreased performance, potentially destabilizing the G&C system [36]. Second, using our reward formulation, the agent has an incentive to deviate from minimizing the LOS rotation rate if it increases the probability of receiving the terminal reward, particularly if the deviation occurs close to the end of an episode. Since minimizing the LOS rotation rate is not an optimal strategy for the case of a maneuvering target (hence the use of augmented proportional navigation [34]), giving the policy the flexibility to deviate from this goal can enhance performance. The third factor is the decreased flight control response time of the integrated system. Specifically, with respect to the longitudinal benchmark, attempting to reduce either the seeker time constant (0.15 s) or noise filter time constant (0.1 s) destabilized the G&C system. Similarly, reducing the FCS time constant below 0.1 s was counterproductive. In contrast, the meta-RL optimized G&C system used a noise filter time constant of 0.01 s. Since the recurrent layer can implement an infinite impulse response filter, it is difficult to determine the overall time constant of the integrated G&C system. However, the time constants of peripheral systems are small, and considering the performance differential we can speculate that the system time constant is much lower than that of the longitudinal benchmark. It is also of interest that the meta-RL optimized G&C system used a maximum fin deflection rate of 80 deg/s, whereas at high altitudes the longitudinal benchmark required a fin deflection rate exceeding 500 deg/s [23].

The design process for meta-RL optimization of an integrated G&C system is more straightforward and robust than traditional approaches. For example, a three loop autopilot is typically designed using a simplified model of the guidance system, with unrealistic assumptions such as a linearized airframe model, constant missile speed, constant altitude, and a body lifting force that is linear in angle of attack [4], and is independent of guidance system optimization. Consequently, when the guidance and flight control systems are simulated together in a high fidelity simulator, FCS gains will likely need to be adjusted, especially in flight regimes where the airframe linearization is inaccurate. In contrast, the meta-RL optimization framework can directly optimize an integrated and adaptive G&C system using a high fidelity simulation environment. In practice, the simulator can instantiate a reduced order aerodynamics model [37] built from computational fluid dynamics simulations. Similarly, the simulator can instantiate high fidelity reduced order radome, actuator, and noise models, and the meta-RL optimization framework is compatible with simulating hardware in the loop. Thus, the meta-RL optimization framework has the potential to reduce the time and cost required to develop a new missile system, potentially providing zero-shot transfer learning between simulation and flight. Moreover, since the meta-RL optimized policy can adapt to novel conditions not seen during optimization, the framework should make successful flight tests more likely. Finally, there should not be any issues implementing the guidance policy on a flight computer, as although it can take several days to optimize a policy, the deployed policy can be run forward in a few milliseconds, as the forward pass consists of a few multiplications of small matrices.

# V. Conclusion

We created a missile aerodynamic model using slender body and Newtonian impact theory, and developed a six degrees-of-freedom simulator to model a range of air-to-air missile head-on engagement scenarios, with the simulator modeling line of sight stabilization, radome refraction, angle measurement noise, and rate gyro bias. The interception problem was then formulated in the meta reinforcement learning framework, using a reward function that minimizes the line of sight rotation rate while imposing path constraints on load and look angle, and an observation space that includes only observations that can be obtained from sensor outputs with minimal processing. The optimized policy implements an integrated and adaptive guidance and control system, directly mapping navigation system outputs to commanded control surface deflection rates. We found that the optimized guidance and control system is robust to moderate levels of radome refraction and rate gyro scale factor bias, performs well against challenging target maneuvers, can adapt to a large flight envelope, and generalizes well to novel conditions not experienced during optimization, including large aerodynamic coefficient perturbations and flexible body dynamics. We found that our guidance and control system significantly outperformed a longitudinal benchmark implementing proportional navigation and a three loop autopilot. Importantly, our experiments clearly demonstrate that including at least one recurrent layer in the policy network is critical to learning an effective G&C system. Future work will model the entire missile trajectory including the rocket burn phase, and attempt to improve performance for the case of an infrared seeker, where measurement of range and range rate is not possible.